\documentclass[preprint,aps,showpacs,showkeys,altaffilletter,preprintnumbers,nofootinbib,amsmath,amssymb]{revtex4}
\usepackage{graphicx,bm,longtable,array,multirow}

\newcommand{\numubar}{\overline{\nu}_\mu}
\newcommand{\numu}{\nu_\mu}

\begin{document}

\preprint{PUCP-FIS-AE-001}

\title{On the Parameters determining the Neutrino Flux \\from observed Active Galactic Nuclei}

\author{J. L. Bazo}
 \email{jlbazo@fisica.pucp.edu.pe}
\author{A. M. Gago}%
 \email{agago@fisica.pucp.edu.pe}
\affiliation{%
Secci\'on F\'{\i}sica, Departamento de Ciencias, Pontificia
Universidad Cat\'{o}lica del Per\'{u}, Apartado 1761, Lima,
Per\'{u}}

\date{\today}

%%%%%%%%%%%%%%%%%%%%%%%%%%%%%%%%%%%%%%%%%%%%%%%%%%%%%%%%%%%%%%%%%
%    Abstract
%%%%%%%%%%%%%%%%%%%%%%%%%%%%%%%%%%%%%%%%%%%%%%%%%%%%%%%%%%%%%%%%%

\begin{abstract}
Extrapolating from a sample of 39 AGNs, we examine the impact on the
total number of high energy $\numu$ induced events (PeV-EeV)
expected in IceCube (a $1Km^3$ ``neutrino telescope''), due to
variations in different parameters involved in the neutrino flux,
such as the emission region geometry, the estimation models and
distributions of the Doppler factor and the variability time. This
work has been done taking into account different limits of the
extragalactic neutrino flux.

Among our conclusions, we find, in the case of the largest
variability time, that the cylindrical geometry hypothesis for the
emission region, produce a separation of 3$\sigma$ in the total
number of events relative to the spherical hypothesis. In addition,
for similar choices of the burst time, spherical geometry and for
the upper neutrino flux bound, we obtain a separation of $2.5
\sigma$ in the total number of events, for some of the Doppler
factor estimations. These differences are undistinguishable for
other input values.
\end{abstract}


\maketitle

%%%%%%%%%%%%%%%%%%%%%%%%%%%%%%%%%%%%%%%%%%%%%%%%%%%%%%%%%%%%%%%%%%%%%%%
\section{\label{sec:intro} Introduction}
%%%%%%%%%%%%%%%%%%%%%%%%%%%%%%%%%%%%%%%%%%%%%%%%%%%%%%%%%%%%%%%%%%%%%%%

Nowadays, there is a special interest in studying different cosmic
rays sources, for instance, Active Galactic Nuclei (AGN) \cite{
Halzen, Atoyan, Becker}, Gamma Ray Bursts (GRB) \cite{Alvarez-Muñiz,
Alvarez-Muñiz2, Dermer, Halzen4}, Dark Matter and others
\cite{Halzen3}. They all help us to understand the structure and
formation of our Universe. In particular, AGNs are very attractive
to investigate, given that they are one of the most powerful
continuous sources of the highest energy particles in the universe.

Among these particles, AGNs produce copiously neutrinos. Due to their weak interaction with matter, neutrinos keep their flux unaltered
in their travel from the AGN to Earth. This feature makes them an
ideal tool for exploring physical processes within the AGNs.

At present, our knowledge about the parameters involved in the
calculation of the extragalactic neutrino flux is not so accurate.
In fact, there is not an absolute agreement about some aspects
related to these parameters. This is the case of the model used to
estimate the Doppler boosting factor (e.g. Equipartition
\cite{G97,Güijosa}, Inverse Compton \cite{G97,Güijosa} and
Variability \cite{Lähteenmäki} models) and its distribution
(power-law-like), the shape of the emission region \cite{Protheroe}
(e.g. spherical and cylindrical) and the variability time scale,
which can vary from few minutes to many days. A good determination of the
Doppler factor is a key ingredient to establish the intrinsic photon
luminosity, directly related with the neutrino flux. In the same
way, the shape of the emission region changes the volume considered
in the particle density of target photons that interact with
protons. In addition, the time of burst is inversely proportional to
the neutrino flux, so that, if the variability time decreases, the
neutrino flux increases.

It is straightforward to note that changes in these parameters
affect directly the neutrino flux and, hence, imply modifications in
the expected value of the total number of neutrinos that will be
detected on Earth. Consequently, it would be interesting to know 
the range of variation of these events when different assumptions are made
about the parameters. If the number of events changes in a
noticeable way, these numbers can be used as a tool to test the model
parameters.

In this paper, we study the impact of these variations in the total
number of muon neutrino induced events that could be detected in
IceCube \cite{IceCube}. This detector is one of the so called
``neutrino telescopes'' (e.g. Amanda, Antares, Nestor) and consists
of arrays of photomultipliers, located in a Cherenkov medium (ice),
under construction in the South Pole with a standard volume of
1Km$^3$. We choose ten years of exposure in the energy range of
PeV-EeV. In spite of the fact, that this interval must contain a
small number of events, it is convenient to be studied, considering
the absence of background in this energy region. We model the
neutrino flux based on a catalogue of 39 AGNs and extrapolate for a
maximum flux given for different proposed limits, which are 
the Waxman-Bahcall~\cite{Waxman}, the benchmark used by the 
IceCube Collaboration~\cite{IceCube2} and the 
$\gamma$-ray bound~\cite{Kuzmin}.

%%%%%%%%%%%%%%%%%%%%%%%%%%%%%%%%%%%%%%%%%%%%%%%%%%%%%%%%%%%%%%%%%%%%%%%
\section{\label{sec:um} AGNs and the Unified Model}
%%%%%%%%%%%%%%%%%%%%%%%%%%%%%%%%%%%%%%%%%%%%%%%%%%%%%%%%%%%%%%%%%%%%%%%

Active Galactic Nuclei are one of the brightest known objects in the
Universe, with an observed luminosity ranging between
$10^{42}-10^{49}$ erg/s. According to the Unified Model of AGN
\cite{Urry}, they have a central engine composed by an accretion
disk orbiting a rotating super-massive black hole ($10^6$ to
$10^{10}$ solar masses). The AGN is powered by an accretion rate of
matter onto the super-massive hole of a few solar masses per year,
possibly supplied by the rotational energy of the black hole.

The Unified Model also claims other characteristics of the AGNs,
such as: two jets\footnote{Generic AGN models do not necessarily
involve jets.} moving in opposite directions and outwards to the
accretion disk, an optically thick torus and radio lobes.
Furthermore, for this model Seyferts, Quasars, Blazars and Radio
Galaxies, are all AGN, only with different orientations as viewed
from Earth, accretion rates and black hole masses. In this work, we
assume this model including in our test sample all these types of
galaxies.

%%%%%%%%%%%%%%%%%%%%%%%%%%%%%%%%%%%%%%%%%%%%%%%%%%%%%%%%%%%%%%%%%%%%%%%
\section{\label{sec:neutrinos}Neutrino flux from AGN}
%%%%%%%%%%%%%%%%%%%%%%%%%%%%%%%%%%%%%%%%%%%%%%%%%%%%%%%%%%%%%%%%%%%%%%%

We will consider as our working hypothesis \cite{Halzen1} that
within inner regions of AGN jets, protons are highly accelerated by
sheet-like shock waves (blobs, knots or bursts) to energies up to
the range of PeV-EeV, according to Fermi's stochastic acceleration.
These protons should then interact with ambient photons emitted by
the accretion disk from the thermal blackbody radiation (UV bump or
blue bump, between $7\times10^{14}$Hz to $3\times10^{15}$Hz),
leading to the $\Delta^+$(1232) resonance and followed by the decays
(pion from photo-production) $n\pi^{+}\rightarrow\mu^{+}
\numu\rightarrow\numubar e^{+}\nu_{e}$. Thus, through this cascade,
a neutrino flux is produced.

In order to calculate the observed neutrino flux produced in each
AGN, we begin with the following generic useful relation
\begin{equation}\label{basic}
\int^{E_{\nu}^{max}}_{E_{\nu}^{min}}E_{\nu}\frac{d\Phi_{\nu}}{dE_{\nu}}dE_{\nu}=\frac{L_{\nu}}{4\pi
d_L^2}~,
\end{equation}
where the integration limits depend on the range of energies chosen
to observe (in our case PeV-EeV), $L_{\nu}$ is the observed neutrino
luminosity and $d_L$ is the luminosity distance to the AGN, defined
as
\begin{equation}\label{lumdist} d_L=d_m (1+z)~,\end{equation}
with z the redshift of the AGN and $d_m$ the proper motion distance
(see Eq. (\ref{dist}) in App \ref{sec:data}).

The energy spectrum, $\frac{d\Phi_{\nu}}{dE_{\nu}}$, will be assumed
to be proportional to
\begin{equation}\label{spectrum}
\frac{d\Phi_{\nu}}{dE_{\nu}}\propto E^{-p}_{\nu}\exp
\left(-\frac{E_{\nu}}{E_{cut}^{AGN}}\right)~.
\end{equation}
The typical power law energy dependence of the extragalactic
neutrino flux is a resemblance of the proton spectrum, which is
supposed to be accelerated by the Fermi's shock mechanism. The slope
of the power law, $p$, is usually taken to be 2, the generic
$E^{-2}_{\nu}$. However, it can be considered in the range
$1.3-2.7$. The spectrum also includes an exponential cutoff
depending on the AGN´s maximum attainable energy, $E_{cut}^{AGN}$
(we will assume a general cutoff of $10^{10}$ GeV).

Therefore, the exact neutrino spectrum formula is
\begin{equation}\label{basicflux}
\frac{d\Phi_{\nu}}{dE}=\frac{L_{\nu}}{4\pi d_L^2 }
\left(\int^{E_{\nu}^{max}}_{E_{\nu}^{min}}E_{\nu}E^{-p}_{\nu}\exp\left(-\frac{E_{\nu}}{E_{cut}^{AGN}}\right)dE_{\nu}\right)^{-1}
 E^{-p}_{\nu}\exp\left(-\frac{E_{\nu}}{E_{cut}^{AGN}}\right)~.
 \end{equation}

We focus our attention on the observable quantities that compose the
neutrino luminosity. Then, to unfold this luminosity in terms of
these observables, we start from the definition of luminosity
\begin{equation}\label{deflum} L_{\nu}=\frac{N_{\nu} \left<E_{\nu}\right>}{\Delta t_{obs}} ~,\end{equation}
where $N_{\nu}$ is the number of produced neutrinos,
$\left<E_{\nu}\right>$ the mean neutrino energy and $\Delta t_{obs}$
the observed variability time of the emission region. Now, we give
$N_{\nu}$ from the neutrino production sequence explained at the
beginning of the section
\begin{equation}\label{nunumber} N_{\nu}= (br_{\Delta\rightarrow n\pi^{+}})(
br_{\pi^{+}\rightarrow\numu})N_{\Delta}~,
  \end{equation}
where ``$br$'' stands for the branching ratio of the desired decay
channel. This gives approximately  $N_{\nu}\approx
N_{\pi^{+}}\approx\frac{1}{2}N_{\Delta}$.

It is useful to express the neutrino energy in terms of the energy
of the parent proton by
\begin{equation}\label{nuenerg} \left<E_{\nu}\right>=
\left<x_{p\rightarrow\pi^{+}}\right>\left<x_{\pi\rightarrow\nu_{\mu}}\right>\left<E_{p}\right>~,\end{equation}
where $\left<x\right>$ represents the average fraction of energy
transferred from the parent particle to the product, which
numerically gives $E_{\nu}\approx \frac{1}{4}E_{\mu}\approx
\frac{1}{4}(\frac{1}{5}E_{p})$.

On the other hand, to find the number of produced $\Delta$'s  we use

\begin{equation}\label{deltanum} N_{\Delta}= \Phi_{p}  N_{\gamma}
\sigma_{p\gamma\rightarrow\Delta}\Delta t_{obs}~,\end{equation} with
$\Phi_{p}$ the initial flux of highly accelerated protons,
$N_{\gamma}$ the number of ambient photons from the UV bump and
$\sigma_{p\pi\rightarrow \Delta}$=$10^{-32} m^2$ the delta
photo-production cross section. $\Phi_{p}$ and $N_{\gamma}$ are
given by
\begin{equation}\label{fluxprot} \Phi_{p}= \frac{L_p R}{\left<E_{p}\right> V}
\quad, \quad N_{\gamma}=\frac{L_{\gamma}\Delta
t_{obs}}{\left<E_{\gamma}\right>~,}
\end{equation}
where $L_p$ and $L_\gamma$ are the proton and photon observed
luminosities, respectively, $\left<E_{\gamma}\right>$ is the mean
photon energy (in the present work we take it to be 10 eV), $R$ is
the observed size\footnote{In the case of a spherical or cylindrical
emission region it corresponds to its radius.} of the emission
region and $V$ the observed volume of the emission region. It must
be noted that the size of the emission region and, consequently, its
volume depend on four parameters: the Doppler factor ($\delta$), the
variability time scale ($\Delta t_{obs}$), the redshift (z) and the
angle between the velocity vector of the knot and the line of sight
($\theta$). The detailed description of $R$ and the volume can be
found in Sec. \ref{subsec:geom}.

Replacing Eqs. (\ref{nunumber} - \ref{fluxprot}) into Eq.
(\ref{deflum}), we obtain
\begin{equation}\label{complexlum}  L_{\nu}=\frac{(br_{\pi^{+}\rightarrow\numu})( br_{\Delta\rightarrow n\pi^{+}})
    \left<x_{p\rightarrow\pi^{+}}\right>
    \left<x_{\pi\rightarrow\nu_{\mu}}\right> L_p L_{\gamma}\Delta t_{obs} R \sigma_{p\gamma\rightarrow\Delta}}{\left<E_{\gamma}\right>V}~.
\end{equation}

To write Eq. (\ref{complexlum}) in a more compact notation, we use
the optical depth, $\tau$, defined as
\begin{equation}\label{opt}
    \tau=\frac{R}{\lambda_{p\gamma \rightarrow \Delta}} = R ~
n_{\gamma}\sigma_{p\gamma\rightarrow\Delta}
    = R \frac{L_{\gamma} \Delta t_{obs}}{V \left<E_{\gamma}\right> }\sigma_{p\gamma\rightarrow\Delta}~,\end{equation}
where $\lambda_{p\gamma \rightarrow \Delta}$ is the proton
interaction length, $R$, the observed size, which should be
understood as the linear path traveled by the proton in the
direction of movement through the ambient photons and we write K as
\begin{equation}\label{K} K= (br_{\pi^{+}\rightarrow\numu})( br_{\Delta\rightarrow n\pi^{+}})
    \left<x_{p\rightarrow\pi^{+}}\right>
    \left<x_{\pi\rightarrow\nu_{\mu}}\right>\approx0.024 ~,\end{equation}
where we have used $br_{\Delta\rightarrow\pi}=0.5$,
$br_{\pi\rightarrow\nu}=0.9998$,
$\left<x_{p\rightarrow\pi}\right>=0.2$ and
$\left<x_{\pi\rightarrow\nu}\right>=\frac{1}{4}$.

Since $\tau$ is a Lorentz invariant, we can go from the observer's
frame to the frame attached to the knot, so that we can use a simple
geometry (see Sec. \ref{subsec:geom}) to describe the emission
region, leaving $\tau$ as follows
\begin{equation}\label{opt2}
    \tau=\tau'=
     \frac{R'(\delta,\Delta t_{obs},z,\theta)}{V'(\delta,\Delta t_{obs},z,\theta)}
    \frac{L'_{\gamma}\Delta t'}{\left<E'_{\gamma}\right>} \sigma_{p\gamma\rightarrow\Delta}=\frac{R'(\delta,\Delta t_{obs},z,\theta)}{V'(\delta,\Delta t_{obs},z,\theta)}
    \frac{L_{\gamma}\Delta t_{obs}}{\left<E_{\gamma}\right>} \sigma_{p\gamma\rightarrow\Delta}~.\end{equation}
where primed quantities refer to a quantity measured in the jet and
recalling that $ \frac{L_{\gamma}\Delta
t_{obs}}{\left<E_{\gamma}\right>}$ is itself a Lorentz invariant.
This now gives for Eq. (\ref{complexlum})
\begin{equation}\label{totflux} L_{\nu}= K ~ \tau ~L_p ~.\end{equation}

However, not all $L_p$ will be observed because part of it would
have already interacted with the ambient photons to produce the
$\Delta$'s. Therefore, the intrinsic luminosity will be $e^{\tau}$
times the observed luminosity, $L_p=e^{\tau}L_{p_{obs}}$. In our
analysis we take ${L_{p_{obs}}}$, which is still unknown, as 10\% of
the photon total luminosity \cite{Halzen1}. In addition, we must
take into account the possible absorption of $\pi^+$ into the source
when the medium is optically thick and therefore will not give time
for the $\pi^+$ to decay into neutrinos. In this case, we use the
substitution suggested by Ref. \cite{Halzen1}, changing $\tau$ by
$(1-e^{-\tau})$. Thus, the factor $(1-e^{-\tau})e^{(1-e^{-\tau})}$
accounts for the absorption of p and $\pi^+$ in the source. Finally
the $\numu$ and $\numubar$ differential flux emitted by an
individual AGN source can be obtained, using Eq. (\ref{basicflux}),
with
\begin{align}\begin{split}\label{flux} \frac{d\Phi_{\numu+\numubar}}{dE_{\nu}}
   & =  \frac{2 \frac{0.1 K L_{\gamma_{obs}}~\left[(1-e^{-\tau})e^{(1-e^{-\tau})}\right]}{4 \pi
{d_L}^2}~}{\int^{E_{\nu}^{max}}_{E_{\nu}^{min}}E_{\nu}E^{-p}_{\nu}\exp\left(-\frac{E_{\nu}}{E_{cut}^{AGN}}dE_{\nu}\right)}
 ~E^{-p}_{\nu}\exp\left(-\frac{E_{\nu}}{E_{cut}^{AGN}}\right)
    \\ & = ~ F(\delta,V,\Delta t_{obs},z,L_{\gamma_{obs}},\theta,E_{\nu}^{max},E_{\nu}^{min})~ E^{-p}_{\nu}\exp\left(-\frac{E_{\nu}}{E_{cut}^{AGN}}\right)~,\end{split} \end{align}
where the factor 2 takes into account the contributions of $\numu$
and $\numubar$\footnote{The muon antineutrino produced after the
$\mu^+$ decay has almost the same average fraction of energy
transferred from the $\pi^+$,
$\left<x_{\pi\rightarrow\nu_{\mu}}\right>\approx\left<x_{\pi\rightarrow\numubar}\right>$,
because each lepton shares approximately equal energy.}. The
function $F$ includes the parameters that we are going to study
$\delta$, $V$ and $\Delta t_{obs}$ (the calculation of $\theta$ is
indirectly affected by the estimation model of $\delta$).

%%%%%%%%%%%%%%%%%%%%%%%%%%%%%%%%%%%%%%%%%%%%%%%%%%%%%%%%%%%%%%%%%%%%%%%
\subsection{\label{subsec:analyzed} Flux parameters to be analyzed}
%%%%%%%%%%%%%%%%%%%%%%%%%%%%%%%%%%%%%%%%%%%%%%%%%%%%%%%%%%%%%%%%%%%%%%%

In this subsection we describe the different hypothesis in the
neutrino flux parameters studied in this paper.

\subsubsection{Doppler boosting factor}
As we have shown in the section before, the Doppler factor,
$\delta$, is a piece in the flux formula. At present, there are
various models for computing $\delta$, which, in some cases, give
different output values of $\delta$ for the same object. This
disagreement could also imply differences in the flux that would be
interesting to examine. For our work, we consider the following
estimation models of $\delta$:

\begin{itemize}

     \item Equipartition $\delta$ (EQ) \cite{G97,Güijosa}: This is calculated
assuming the equipartition of energy between radiating particles and
magnetic field and is obtained from the ratio of the observed and
the maximum intrinsic brightness temperatures. The maximum intrinsic
temperature is supposed to be the equipartition temperature.

\item Inverse Compton $\delta$ (IC) \cite{G97,Güijosa}: In this case,
the Doppler factor is determined from the comparison of the observed and
predicted X-ray fluxes, assuming that this is caused by inverse
Compton scattering of synchrotron photons off the radiating particles.

\item Variability $\delta$ (Var) \cite{Lähteenmäki}: This is obtained in a similar
way to $\delta_{EQ}$, but is weaker than this, since it takes the
third root of temperatures' ratio. Additionally, it replaces the
observed brightness temperature instead of the variability
brightness temperature, which is estimated from a total flux density
flares associated with VLBI components emerging from the AGN core,
covering nearly 20 years of variations.

    \item Minimum $\delta$ (Min)
\cite{G93}: This model is computed only from the proper motion data
using the minimum $\delta$ value allowed by
\begin{equation}\label{dmin} \delta_{min}=\Gamma_{min}=\sqrt{1+\beta_{app}^{2}} ~,\end{equation}
where $\Gamma$ is the Lorentz factor and $\beta_{app}$ the apparent
transverse velocity in units of c (the speed of light). We are going
to use this model independent estimation as a reference.

\end{itemize}

Due to different proper motion observations of underlying shocks for
a single jet, there should be a rather broad range of Doppler
factors \cite{K04}. Then, it follows that a certain distribution in
$\delta$ must be taken into account when computing the neutrino
flux. According to Ref.\cite{K04} it is consistent with a steep
power law distribution of intrinsic Doppler factors. In our
calculation, we fit with a power law ($\delta^{-\gamma}$) the data
obtained for each type of calculation of $\delta$, as described in
Appendix.~\ref{sec:data}. In Sec.~\ref{sec:event} we explain how the
neutrino flux is averaged over the fitted distribution.

\subsubsection{\label{subsec:geom} Emission Region Geometry}

In the study of the emission region geometry, we consider two
possibilities for its shape in the  proper frame: a spherical and a
cylindrical volume. These models are based on the observed
variability time $\Delta t_{obs}$ \cite{Protheroe}. However, we
should state that $\Delta t_{obs}$ could give, at most, the
estimation of only one of the dimension of this region and therefore
the other two should be inferred somehow from this. In this sense,
the formulas given here are actually not relativistic
transformations (as in App. \ref{trans}), but rather estimations and
should not be misunderstood.

To understand the next formulas, it is appropriate to define a quantity $D=\frac{\delta}{1+z}$ that
includes both the Doppler factor ($\delta$) and the cosmological
effects given by the redshift (z). The
geometries are described as follows:

\begin{itemize}

    \item Spherical geometry: it has two options for the jet-frame source radius $R'$. The first option, that we call {\bf SphereA}, uses the common formula
         \begin{equation}\label{esf} R'= D c \Delta t_{obs}~. \end{equation}

The second option, that we call by {\bf SphereB}, uses a refined
formula for $R'$, that consider the angle to the line of sight,
$\theta$, and the shock velocity, $\beta_{shock}$ (used value
$\beta_{shock}$=0.5). This expression is given by~\cite{Protheroe}
  \begin{equation}\label{esfA}
    R'=\frac{D c \Delta t_{obs}}{\frac{1}{\sqrt{2}}\operatorname{Abs}\left[\frac{1}{\beta_{shock}}-\Gamma D( \cos (\theta)-\beta)\right]+\frac{D
    \sin(\theta)}{\Gamma}}~.
    \end{equation} \\
    \item Cylindrical geometry: with radius $R'$ and jet-frame length $l$. The region is rapidly energized by a plane shock traveling along
    the jet, such that photons are emitted immediately after shock passage from a thin disk-like region downstream of the shock.
    From here on, we assume $l=aR'$, with $a$ variable (used value $a=2$).
    \begin{equation}\label{cil}
    R'=\frac{D c \Delta t_{obs}}{\frac{a \Gamma}{2}\operatorname{Abs}\left[\frac{1}{\beta_{shock}}-\Gamma D( \cos (\theta)-\beta)\right]+ D
    \sin(\theta)}~.
    \end{equation}

\end{itemize}

\subsubsection{\label{subsec:vartime} Variability Time Scale}

The observed fluctuations over the whole spectrum in the AGNs, or
variability time scales of the bursts, are one of the main
characteristics of Active Galaxies \cite{Uttley}. This variability
suggested that the emitting regions were extremely compact, leading
to the hypothesis that AGN were powered by massive black holes.

The time scale varies according to the class of AGN and waveband of
the spectrum. The shortest fluctuation observed in an AGN is
attributed to Mrk421 with a 15 minutes duration. However, there have
been variability on time scales of months to years \cite{Uttley}. In
the case of blazars, a rapid variability or flaring is common.
Observed short time scales of a day or more in the X-ray band are
characteristic to TeV blazars (e.g. Mrk421, Mrk501, PKS2155-304).
They have even substructures with shorter time scales of $10^3 -
10^5$ sec \cite{Tanihata}. On the other hand, many EGRET sources had
time scales of several days \cite{Halzen1} and the optical waveband
flares are fairly slow.

After this considerations, we choose the variability time
scales ranging from 15 minutes to 10 days.

%%%%%%%%%%%%%%%%%%%%%%%%%%%%%%%%%%%%%%%%%%%%%%%%%%%%%%%%%%%%%%%%%%%%%%%
\section{\label{sec:cat} Our AGN sample}
%%%%%%%%%%%%%%%%%%%%%%%%%%%%%%%%%%%%%%%%%%%%%%%%%%%%%%%%%%%%%%%%%%%%%%%

Although there are more than 876 Blazars, 11777 Seyfert-1 and 48921
Quasi Stellar Objects catalogued by Ver\'{o}n-Cetty
(2003)\cite{Veron}, only few of them contain the parameters
important for our interests. After a thorough search of AGN
catalogs, we have built our own AGN sample from some of them, so
that gathering their data, we included relevant information to our
investigation, such as the proper motion of jet components,
luminosity and different Doppler factor estimations. From the
combination of Ref.\cite{K04}, the latest and widest AGN survey
available on proper motion observations, with Ref.\cite{K98}, which
includes luminosities, and Ref.\cite{Güijosa,Lähteenmäki},
containing Doppler factor estimations, we have selected only the
objects that have information on all parameters, remaining a sample
of 39 AGN. In order to get a proper motion distribution for the
AGN's jets we add more data in the proper motion observations from
earlier surveys \cite{C88,J01,G93,G97,V94,H00}.

We must note that these multiple observations are not from the same
epochs, which might introduce some uncertainties. However, we think
that it is good enough for our purposes.

\subsection{Description of the sample}

Our total sample of 39 AGN, is divided into tree sub-classes: 10 BL
Lacs (blazars) (see Table \ref{tableblazar}), 27 Core-dominated
Quasars (see Table \ref{tablequasars}) and 2 Radio Galaxies (see
Table \ref{tableradiogalaxies}). Tables
[\ref{tableblazar},\ref{tablequasars},\ref{tableradiogalaxies}] show
the observational data that describes the properties of each source
(i.e. AGN model parameters). The data is presented as follows:

\begin{itemize}\addtolength{\itemsep}{-0.6\baselineskip}
\item Column(1) International Astronomical Unit source designation [Name B1950].
\item Column(2) Alternative source Name.
\item Column(3) Redshift.
\item Column(4) Monochromatic luminosity at 15 GHz, from \cite{K98}, completed with \cite{Z02}.
\item Column(5) Component identifier of the knot.
\item Column(6) Angular radial speed: proper motion.
\item Column(7) Equipartition Doppler factor, from \cite{Güijosa}.
\item Column(8) Inverse Compton Doppler factor, from \cite{Güijosa}.
\item Column(9) Variability Doppler factor, from \cite{Lähteenmäki}.
\item Column(10) Reference for proper motion.
\end{itemize}

A comment is in order, in our calculations of the neutrino flux, we
need to use the UV luminosity $L_{\gamma_{obs}}$ in Eq.~(\ref{flux})
(see Sec.~\ref{sec:neutrinos}). For that reason we extrapolate the
luminosity given in this sample, which is in the radio component of
the spectrum, to find the UV luminosity. This estimate has been done
in the following way: from the luminosity fractions (for different
parts of the electromagnetic spectrum) of the set of AGNs presented
in Ref.~\cite{Impey}, we compute the average luminosity fractions,
such that the radio part represents 0.52\% and the optical-UV region
26\%. Then, with the relation $L_{UV}\approx50L_{rad}$ we do the
extrapolation.

\subsection{Sample characteristics}

To better understand the characteristics of the AGN sample, we
present in this subsection several histograms and comparisons of its
most relevant properties.

\begin{figure}[h]
\begin{center}
  \includegraphics[width=9cm,keepaspectratio=true]{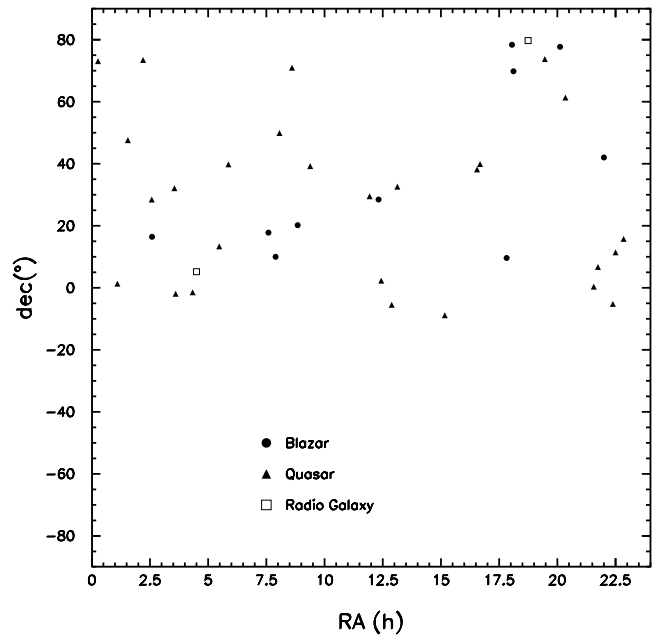}\\
  \caption{Distribution of the AGNs sample given in right ascension and declination angles.}\label{agnplaces}
\end{center}
\end{figure}

We display in Fig.~\ref{agnplaces} the angular distribution, in
right ascension and declination, of our sample's 39 AGNs. Most of
the sources are located in the northern hemisphere due to the
abundance of observatories in that region. Nevertheless, this bias
does not affect our purposes, since we are going to consider the
neutrino flux as isotropic.

\begin{figure}[h]
\begin{center}
  \includegraphics[width=9cm,keepaspectratio=true]{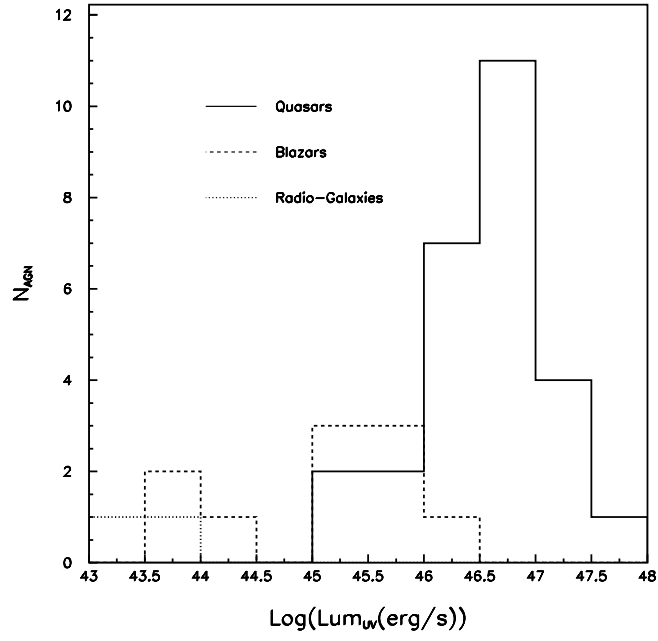}\\
  \caption{Distribution of the observed UV-Luminosity in the sample.}\label{histlum}
\end{center}
\end{figure}

We show in Fig.~\ref{histlum} the extrapolated UV luminosity (in the
observer's frame) distribution of our sample. It can be seen that
Quasars have the highest luminosity, with a mean value of
$7.23\times10^{46}$ erg/sec, while Blazars and Radio-Galaxies have
$4.82\times10^{45}$ erg/sec and $4.50\times10^{45}$ erg/sec,
respectively. These observations are consistent with our
expectations, since Quasars are thought to be the most luminous
objects in the universe.

\begin{figure}[h]
\begin{center}
  \includegraphics[width=9cm,keepaspectratio=true]{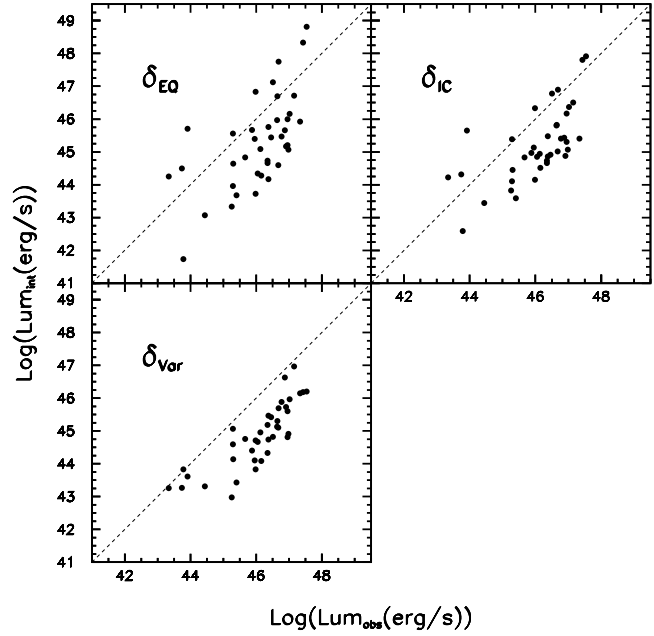}\\
  \caption{Intrinsic versus observed luminosities according to the
Doppler estimation model.}\label{lobsvslint}
\end{center}
\end{figure}

In Fig.~\ref{lobsvslint} we present the extrapolated UV luminosity
versus its intrinsic counterpart, which was obtained using Eq.
(\ref{lum}), for each Doppler factor estimation model. From that
equation we expect that in the case of a $\delta$ slightly greater
than one, the observed luminosity should be higher than the
intrinsic luminosity. This situation is reflected in our sample,
where the observed luminosities for each AGN is higher than their
corresponding intrinsic values, which is represented by the points
below the straight line. This indicates that most AGN jets in the
sample are moving towards Earth and suffer the Doppler boosting.
Only few AGN, in the case of the Equipartition and Inverse Compton
estimation models, have the opposite behavior. This is because, for
these models, some Doppler factor estimates are lower than one. On
the contrary, for the Variability model all Doppler factor estimates
are greater than one, meaning that their luminosities will suffer
the so called Doppler boosting.

\begin{figure}[h]
\begin{center}
  \includegraphics[width=9cm,keepaspectratio=true]{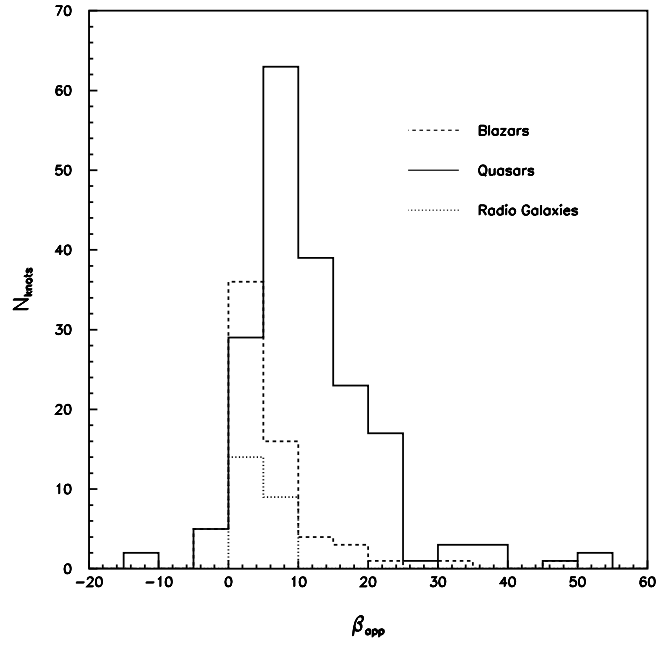}\\
  \caption{Apparent transverse velocity distribution in the sample.}\label{histbeta}
\end{center}
\end{figure}

The apparent transverse velocity distribution for each knot in the
sample is shown in Fig.~\ref{histbeta}. From this plot it is clear
to note that Quasars have higher values of apparent transverse
velocities than Blazars and Radio Galaxies, which is in agreement
with Fig.~\ref{histlum}, due to the relationship between Doppler
factor, which depends on the apparent transverse velocity, and
luminosity. We also notice here that some of the AGN jets are moving
away from Earth, which match with our observations in
Fig.~\ref{lobsvslint}.

The distribution of the mean angle to the line of sight, calculated
using the method described in Appendix \ref{sec:data}, is given in
Fig.~\ref{histangle} for different Doppler factor estimation models.
As we could have supposed from the previous plots, we see that the
majority of the AGN jets are pointing almost directly towards us.
This fact does not entirely reflects the assumptions of the Unified
Model, which predicts, in average, greater angles for Quasars and
Radio-Galaxies, than we have in our histograms. Meanwhile, Blazars
are compatible with small angles, which is expected, since their
jets are pointing almost directly to us. The mean $\theta$ values
found for the whole sample are for
$\theta_{EQ}\approx5.1$\textordmasculine,
$\theta_{IC}\approx9.4$\textordmasculine,
$\theta_{Var}\approx7.6$\textordmasculine, and
$\theta_{min}\approx13.4$\textordmasculine.

\begin{figure}[h]
\begin{center}
  \includegraphics[width=9cm,keepaspectratio=true]{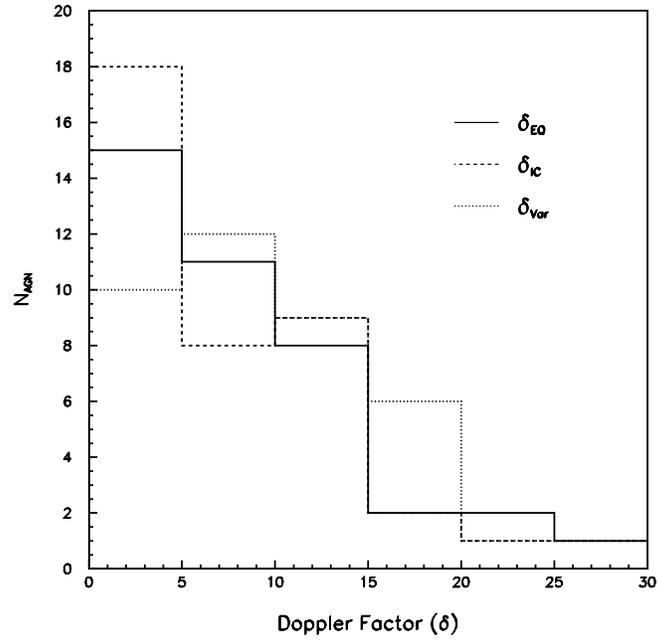}\\
  \caption{Doppler factor model distribution in the sample.}\label{histdoppler}
\end{center}
\end{figure}

In Fig.~\ref{histdoppler} we plot the distribution of the Doppler
factor for each estimation model. We note that the Equipartition and
Inverse Compton cases give lower $\delta$ values than the
Variability case. Their mean values are $<\delta_{EQ}>=8.00$,
$<\delta_{IC}>=7.65$ and $<\delta_{Var}>=9.61$ and they range
between $\delta_{EQ}=[0.14,29]$, $\delta_{IC}=[0.15,27]$ and
$\delta_{Var}=[0.98,26.21]$. We found that these histograms can be
fitted with the expected power law distribution. After applying the
distribution discussed in Appendix \ref{sec:data} we obtain:
$<\delta_{EQ}>=8.11$, $<\delta_{IC}>=6.54$, $<\delta_{Var}>=8.49$
and $<\delta_{min}>=7.04$. These new values correspond to a power
law distribution with negative slope ($\delta^{-\gamma}$) with
exponent $\gamma_{EQ}=0.38$, $\gamma_{IC}=0.51$,
$\gamma_{Var}=0.53$, and $\gamma_{min}=0.55$.

\begin{figure}[h]
\begin{center}
  \includegraphics[width=9cm,keepaspectratio=true]{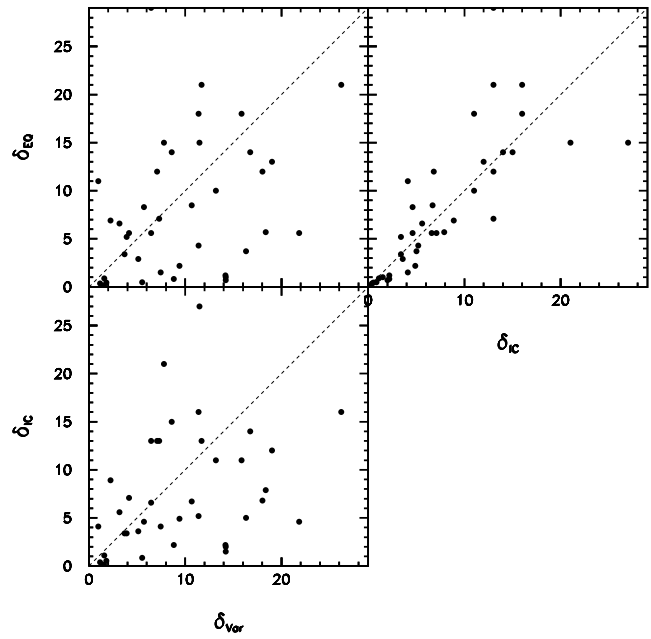}\\
  \caption{Comparison of the different types of Doppler factors 
for the sample.}\label{comparedelta}
\end{center}
\end{figure}

Finally, we compare in Fig.~\ref{comparedelta}, using
two-dimensional plots, the different types of estimations for the
Doppler factor, $\delta_{EQ}$, $\delta_{IC}$ and $\delta_{Var}$. We
find that $\delta_{EQ}$ and $\delta_{IC}$ are strongly correlated,
in contrast, $\delta_{Var}$ is uncorrelated with any of them.

%%%%%%%%%%%%%%%%%%%%%%%%%%%%%%%%%%%%%%%%%%%%%%%%%%%%%%%%%%%%%%%%%%%%%%%
\section{\label{sec:event} \boldmath {$\nu_{\mu}$}-induced event number calculation}
%%%%%%%%%%%%%%%%%%%%%%%%%%%%%%%%%%%%%%%%%%%%%%%%%%%%%%%%%%%%%%%%%%%%%%%

We calculate the number of AGN-neutrino induced events observed on
Earth within the context of the proposed IceCube \cite{IceCube}
detector. As we are interested in astrophysical point sources (i.e.
AGNs), we use muon neutrinos because they can achieve the highest
angular resolution (0.7\textordmasculine) due to the large muon
track, which allows to reconstruct their direction \cite{Karle}.
Moreover, $\numu$-induced events have a higher effective
volume~\cite{Karle} than other neutrino flavors. We restrict our
search to the highest energy events in the PeV-TeV range, within
IceCube's sensitivity~\cite{IceCube}, thus we avoid the background
signals from atmospheric neutrinos. Then, the number of
$\numu+\numubar$ induced muon events is calculated by \cite{Halzen3}
\begin{equation}\label{num} N_{\mu}= T \int d\Omega
\int^{10^{18}~\mbox{\scriptsize{eV}}}_{10^{15}~\mbox{\scriptsize{eV}}}dE_{\numu}
\left< \frac{d\phi_{\nu_{\mu}}}{dE_{\numu}}\left(E_{\numu}\right)
\right>_{tot}A_{eff}\left(E_{\numu}\right)P_{surv}\left(E_{\numu},\theta\right)P_{\numu\rightarrow
\mu}\left(E_{\numu},\theta,E_{\mu}^{thr}\right)~,
\end{equation}
where $T$ is the exposure time (we use 10 years in our calculations)
and $A_{eff}$ is the effective area of the detector, taken from
Ref.~\cite{IceCube2}. Since, for this energy range, IceCube is
capable of observing both hemispheres, we integrate over the whole
celestial sphere. Notice that in this expression, $\theta$
corresponds to the zenith angle ($\theta=0$ is pointing to the South
Pole) and should not be confused with the angle to the line of sight
used to describe the neutrino flux in the previous sections.

The total isotropic flux of AGN neutrinos summed over all sources is
defined as
\begin{align}\begin{split}\label{num2} \left< \frac{d\phi_{\nu_{\mu}}}{dE_{\numu}} \right>_{tot}
    = ~ C_n &
\sum_{j=1}^{39} \left(
\int^{{\delta_{max}}_j}_{{\delta_{min}}_j}d\delta ~ pdf
\left(\delta\right)_{j,mod} F_j\left(\delta,V,\Delta t,z_{j},
{L_{\gamma_{obs}}}_{j},\left<\theta\right>_{(j)mod},10^{18}~\mbox{\scriptsize{eV}},10^{15}~\mbox{\scriptsize{eV}}\right)\right)
    \\ & \times E^{-p}_{\nu}\exp\left(-\frac{E_{\nu}}{E_{cut}^{AGN}}\right) \\=C_n \left<\phi\right>_{tot} &\times E^{-p}_{\nu}\exp\left(-\frac{E_{\nu}}{E_{cut}^{AGN}}\right) ~,\end{split} \end{align}
where $\left<\phi\right>_{tot}$ represents the sum of the terms
$F_j$ (see Eq.~(\ref{flux})), after averaged over the probability
density function ($pdf$) in $\delta$, defined for each $j$ source of
the total $39$ AGN sources of our sample. The probability density
function varies for each estimation model ($mod$) and is given by
\begin{equation}\label{pdf} pdf(\delta)_{j,mod}=\frac{\delta^{-\gamma}}{\int^{{\delta_{max}}_j}_{{\delta_{min}}_j} \delta^{-\gamma}~d\delta}~,
\end{equation}
where $\gamma$ corresponds to each Doppler factor estimation model
($\gamma_{EQ}=0.49$, $\gamma_{IC}=0.58$, $\gamma_{Var}=0.55$,
$\gamma_{min}=0.76$). The integration limits, ${\delta_{max}}_j$ and
${\delta_{min}}_j$, are chosen for a single AGN source, taking the
maximum (${\delta_{max}}_j$) and minimum (${\delta_{min}}_j$) of its
corresponding set \footnote{When there is only one value for an AGN
source, we assume an average range.} of Doppler factors (these
limits depend on the $\delta$ estimation model). The normalization
constant $C_n$ in Eq. (\ref{num2}) allows us to extrapolate the
total $\numu$ flux of our small sample to the maximum possible. This
constant is given by
\begin{equation}\label{Cn} C_n=\frac{\phi_{bound}}{\left<\phi\right>^{max}_{tot}}~,
\end{equation}
where $\phi_{bound}$ stands for the three different limits that we
have chosen: the Waxman-Bahcall limit $3\times 10^{-8}$
GeV$^{-1}$cm$^{-2}$s$^{-1}$sr$^{-1}$ \cite{Waxman}, the benchmark
used by the IceCube Collab. $1\times 10^{-7}$
GeV$^{-1}$cm$^{-2}$s$^{-1}$sr$^{-1}$ \cite{IceCube2} and the
$\gamma$-ray bound $6\times 10^{-7}$
GeV$^{-1}$cm$^{-2}$s$^{-1}$sr$^{-1}$ \cite{Kuzmin}. We must note
that the given bounds are still below the limits set by recent
experiments and above IceCube's sensitivity \cite{IceCube2}. The
limit for energies between 1 PeV and 3 EeV from the latest AMANDA
results \cite{AMANDA} derived with a 90\% C.L. is
$E^2\Phi_{\nu}(E)<0.99 \times 10^{-6}$
GeV$^{-1}$cm$^{-2}$s$^{-1}$sr$^{-1}$  and the AMANDA-II-2000 data is
still being processed. $\left<\phi\right>^{max}_{tot}$ at 1 GeV, is
the maximum value among the set of values of
$\left<\phi\right>_{tot}$ obtained for all the variations performed
in our work (different types of $\delta$ estimation, class of
geometry emission region and variability time of the burst). It is
relevant to mention that there is a possibility that these sources
were optically thick enough to absorb most of the protons, only
allowing neutrinos to scape, which would indeed exceed the
conservative WB bound.

Continuing with the description of Eq. (\ref{num}) we define
$P_{surv}\left(E_{\numu},\theta\right)$, the survival probability of
a neutrino crossing the Earth, by
\begin{equation}\label{Psurv}
P_{surv}\left(E_{\numu},\theta\right)\equiv \exp
\left[-\frac{\chi\left(\theta\right)}{L_{int}^{tot}\left(E_{\numu}\right)}
\right]~,
\end{equation}
where $\chi\left(\theta\right)$ is the column density given by
\begin{equation}\label{X}
\chi\left(\theta\right)\equiv
\int_{0}^{l_{\numu}\left(\theta\right)}\rho\left( r\left(
\theta,l\right)\right)dl~,
\end{equation}
with, $l_{\numu}$, the neutrino distance traveled through Earth
calculated using
\begin{equation}\label{lnu}
l_{\numu}\left(\theta\right)=\sqrt{(R_{E}-d)^{2} \cos^2 \left(\theta
\right)+2d R_{E}-d^{2}}-(R_{E}-d)\cos \left(\theta\right)~,
\end{equation}
where $R_{E}$ is the Earth's radius (6378 Km) and $d$ the distance
to the detector (for IceCube it is 1.4 Km).

In Eq. (\ref{X}) we integrate over the neutrino path crossing the
Earth using the parameterized Earth's density, $\rho\left(r\right)$,
given by the Preliminary Earth Model (PREM \cite{PREM}) and the
inclusive interaction length, $L_{int}^{tot}$, is given by
\begin{equation}\label{Lint}
L_{int}^{tot}\left(E_{\numu}\right)\equiv \frac{1}{\sigma_{\nu
N}^{tot}\left(E_{\numu}\right)N_{Av}}~,
\end{equation}
where the inclusive cross section, $\sigma_{\nu N}^{tot}$, is taken
from Ref.~\cite{Gandhi}, which includes the charged and neutral
current contributions.

The probability that a muon produced in a $\numu$ charged current
interaction reaches the detector before losing all its energy down
to a value lower than the detector's energy threshold
$E_{\mu}^{thr}$ ($10^3$ GeV for IceCube) is defined by \footnote{For
down-going muons if $R_{\mu}(\theta)>d$, then we chose
$R_{\mu}(\theta)=d$, because the Earth's density traversed by
neutrinos is smaller.}
\begin{equation}\label{Pnumu}
P_{\numu\rightarrow
\mu}\left(E_{\numu},\theta,E_{\mu}^{thr}\right)\equiv
\frac{1}{\sigma_{CC}\left(E_{\numu}\right)}\int_{0}^{1-\frac{E_{\mu}^{thr}}{E_{\numu}}}dy\frac{d\sigma_{cc}}{dy}\left(E_{\numu},y\right)\left(1-\exp
\left[-\frac{R_{\mu}(E_{\numu},\theta,E_{\mu}^{thr})}{L_{int}^{cc}\left(E_{\numu}\right)}
\right]\right)~,
\end{equation}
where the differential charged current cross section,
$\frac{d\sigma_{cc}}{dy}$ is taken from Ref.~\cite{Gandhi2} and the
muon range, $R_{\mu}$, is taken from Ref.~\cite{Lipari}.

We include the effects of
neutrino oscillations \cite{Beacom} by multiplying Eq. (\ref{num})
by 0.5, given that the ratio of the initial to the oscillated muon
neutrino flux, $\phi_{\nu_{\mu}}:\phi^{osc}_{\nu_{\mu}}$, would be
2:1.

It is worth mentioning that there are other scenarios where muons
can lose a significant amount of energy before their decay,
equivalent to considering only a single $\numu$ \cite{Levinson},
otherwise we would have to consider half of the neutrino flux
(without neutrino oscillations), which is different from the one we
assume here.

%%%%%%%%%%%%%%%%%%%%%%%%%%%%%%%%%%%%%%%%%%%%%%%%%%%%%%%%%%%%%%%%%%%%%%%
\section{\label{sec:results} Results}
\begin{figure}[p]
\begin{center}
  \includegraphics[height=17.5cm,keepaspectratio=true]{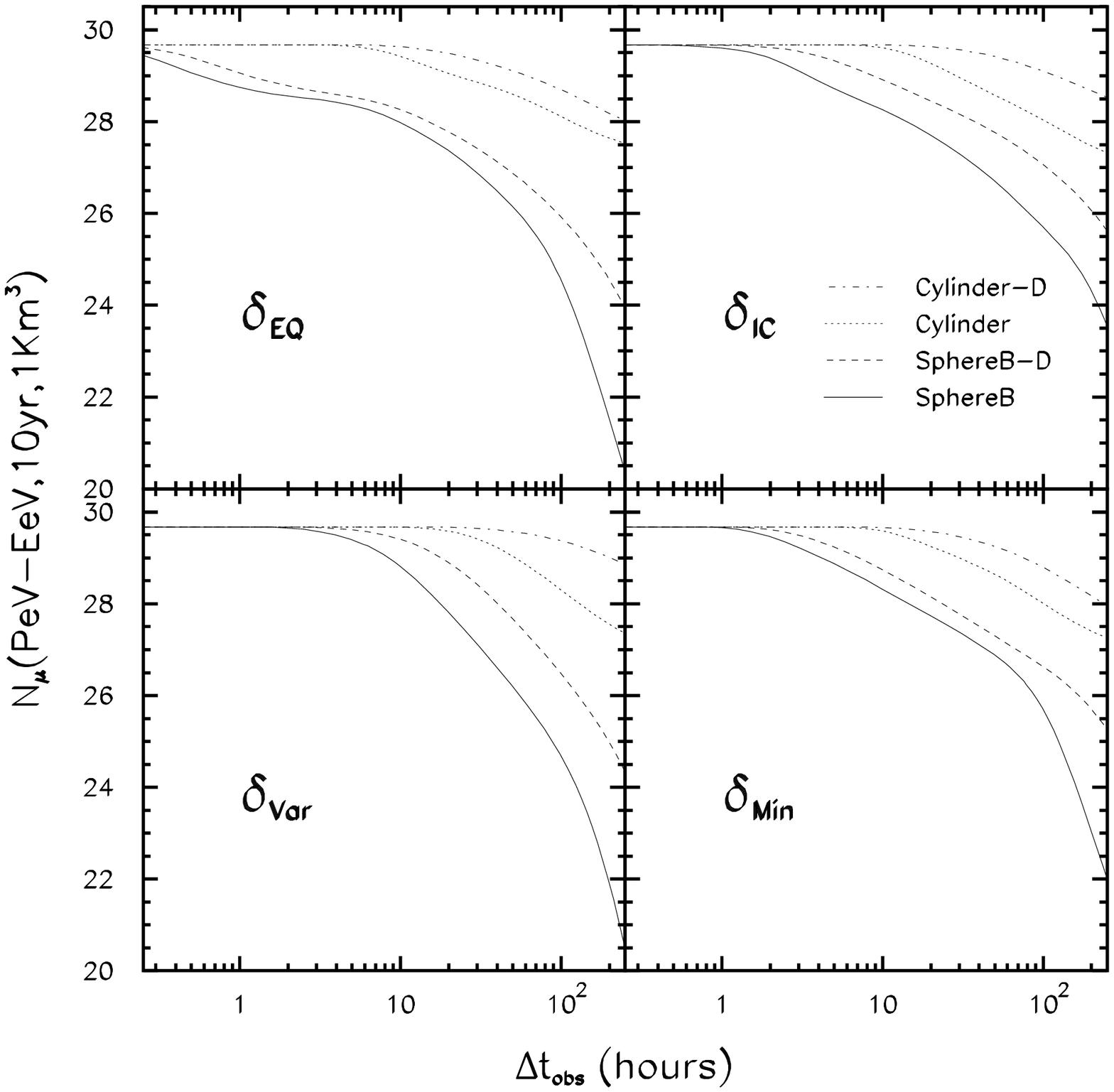}\\
  \caption{Expected total AGN $\numu$-induced events as a function of 
the variability time, for different assumptions 
in the model parameters studied. The maximum flux is 
normalized with the Waxman-Bahcall limit \cite{Waxman} 
and $p=2$ is taken in the energy spectrum. ``$D$'' stands for 
the use of $\delta$ distribution.}
\label{eventvstimewb}
\end{center}
\end{figure}

\begin{figure}[p]
\begin{center}
  \includegraphics[height=17.5cm,keepaspectratio=true]{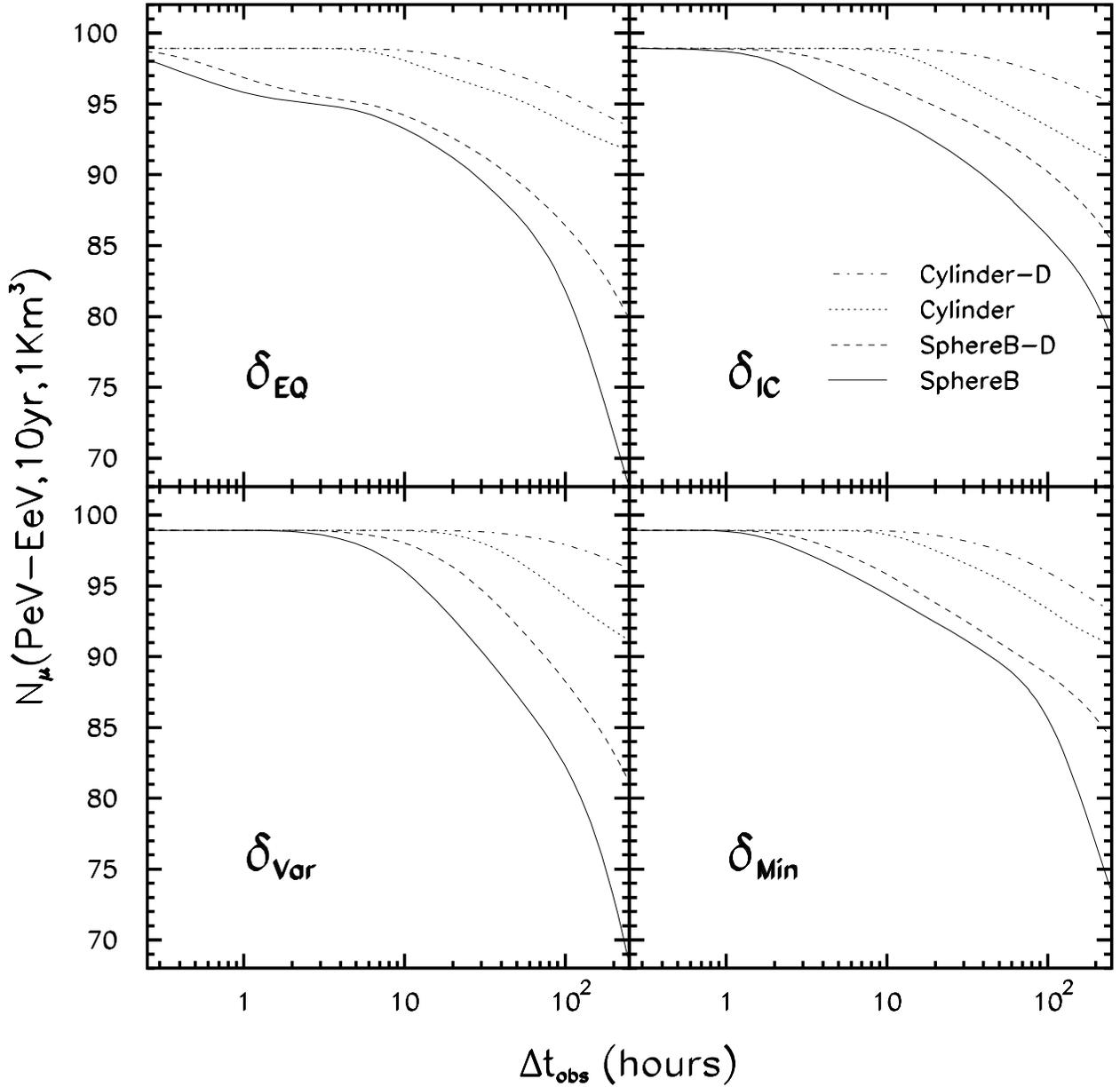}\\
  \caption{Expected total AGN $\numu$-induced events as a function of 
the variability time, for different assumptions 
in the model parameters studied. The maximum flux is 
normalized with the benchmark used by the IceCube Collab. \cite{IceCube2} 
and $p=2$ is taken in the energy spectrum. ``$D$'' stands for 
the use of $\delta$ distribution.} \label{eventvstimebe}
\end{center}
\end{figure}

\begin{figure}[p]
\begin{center}
  \includegraphics[height=17.5cm,keepaspectratio=true]{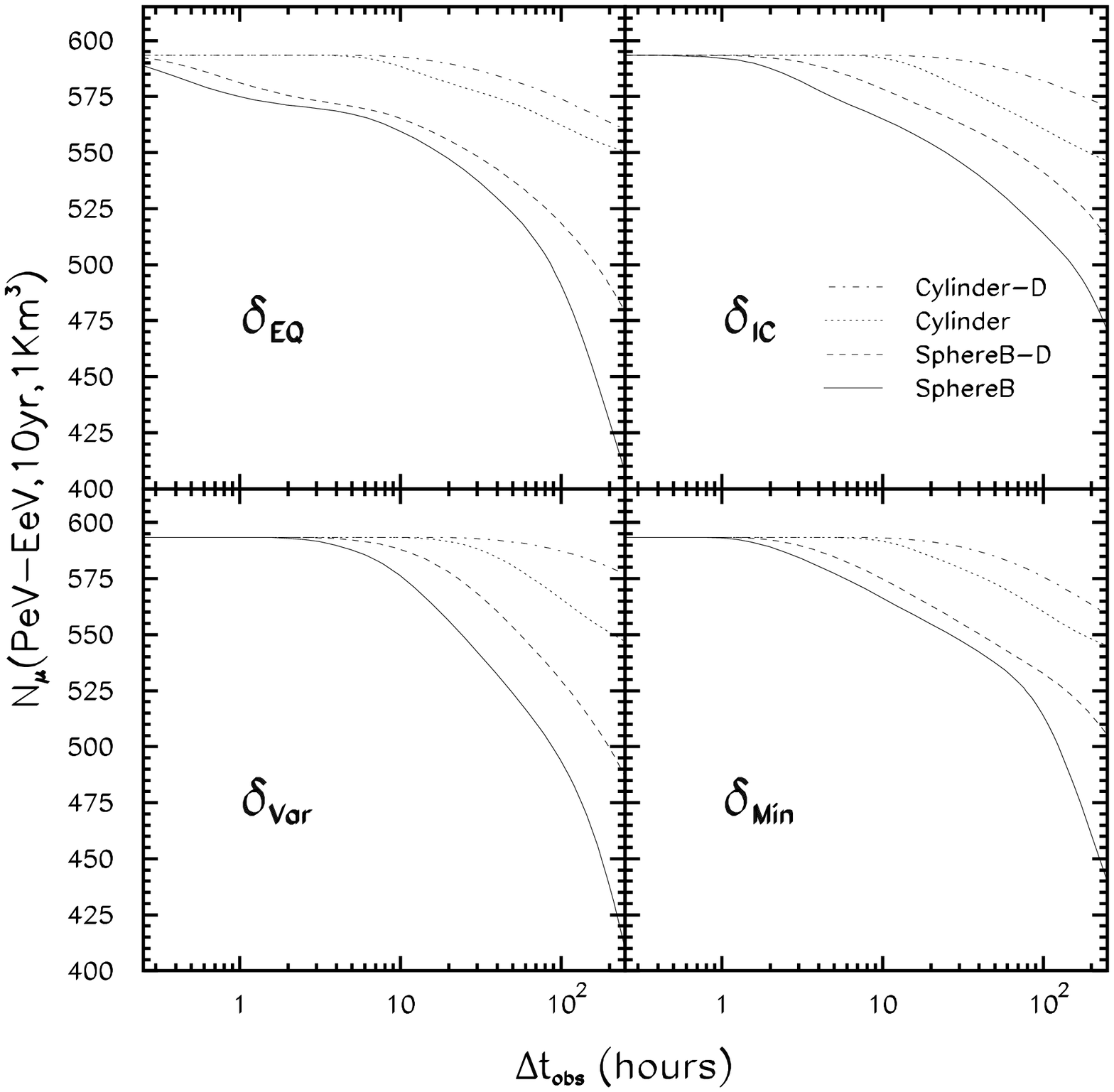}\\
\caption{Expected total AGN $\numu$-induced events as a function of 
the variability time, for different assumptions 
in the model parameters studied. The maximum flux is 
normalized with the $\gamma$-ray bound
\cite{Kuzmin} and $p=2$ is taken in the energy spectrum. ``$D$'' stands for 
the use of $\delta$ distribution.} \label{eventvstimegray}
\end{center}
\end{figure}

%%%%%%%%%%%%%%%%%%%%%%%%%%%%%%%%%%%%%%%%%%%%%%%%%%%%%%%%%%%%%%%%%%%%%%%
We present in Fig.~\ref{eventvstimewb}, Fig.~\ref{eventvstimebe} and
Fig.~\ref{eventvstimegray}, an analysis of the dependence of the
expected total number of induced $\nu_{\mu}$ and $\numubar$ events
versus the variability time (ranging from 15 minutes to 10 days) for
different Doppler factor estimation models, emission region
geometries, and for ten years of exposure in IceCube. We use the
generic spectrum  $E_{\nu}^{-2}$ and normalize the figures with the
WB limit (Fig.~\ref{eventvstimewb}), the IceCube Collab. benchmark
(Fig.~\ref{eventvstimebe}) and the $\gamma$-ray bound
(Fig.~\ref{eventvstimegray}). In each figure we show four plots
related to the different type of $\delta$, where we display four
curves corresponding to the following hypotheses in the geometry of
the emission region: SphereB, Cylinder, SphereB-D and Cylinder-D
(the last two are averaged over the $\delta$ distribution).

In general, we observe that the tendencies of the three figures are
the same and they only differ in the scale of number of events. The
behaviour in $\Delta t_{obs}$ of the curves within each figure, can
be understood from the factor $(1-e^{-\tau})e^{(1-e^{-\tau})}$ in
Eq.~(\ref{flux}), with $\tau \propto 1/\Delta t_{obs}$. Also, using
this factor, we can comprehend two important features of our plots.
First, we see why all the graphs converge to the same number of
events (given for the flux limit) for an small $\Delta t_{obs}$,
which implies that for this time we are not able to distinguish
among the different model parameters under study. Second, we
observe the hierarchy of the curves, where the number of events is
higher for the cylinder than for the sphere geometry. This is not
strange since the spherical hypothesis gives a greater volume than
the cylindrical one, which implies fewer number of events given that
$\tau\propto1/V$. Furthermore, comparing the number of events for
the same geometry, we obtain slightly higher values in the case
where we average over the Doppler factor distribution; this is
because the distribution includes lower values of $\delta$, which
gives, in general, a greater number of events.

In order to quantify the differences observed in our plots, at the
largest variability time, in the number of events, for the different
parameter hypotheses studied (e.g. type of geometries and Doppler
factor calculations), we have defined the following estimator:
\begin{equation}
N_{\sigma}=\frac{\left|
N^{limit}_{hyp_{(i)}}-N^{limit}_{hyp_{(j)}}\right|}{\sqrt{N^{limit}_{hyp_{(j)}}}}~,
\label{estimator}
\end{equation}
which gives us these discrepancies in terms of the number of
$\sigma$'s, with $N^{limit}_{hyp_{(i)}}$ the total number of
expected neutrinos, where $limit$ stands for the bound used in the
extragalactic neutrino flux and $hyp_{(i)}$ refers to the various
hypotheses compared.

Our first analysis has the purpose to evaluate how large the
differences in the number of events are for the spherical and
cylindrical geometry. This is shown in Table~\ref{sigma}, where we
have substituted in Eq.~(\ref{estimator}) $hyp_{(j)}=$
SphereB(SphereB-D) and $hyp_{(i)}=$ Cylinder(Cylinder-D), evaluated
for each type of Doppler calculation. From this Table, we can see
that, in the case of the WB limit, the differences go from
0.75$\sigma$ to 1.53$\sigma$, when we do not include the $\delta$
distribution. Meanwhile, when the $\delta$ distribution is
considered, these differences are even smaller going from
0.53$\sigma$ to 0.89$\sigma$. For the IceCube benchmark, the
differences are significant and range between 1.36-2.79$\sigma$ and
1.02-1.62$\sigma$ for the cases without and with $\delta$
distribution, respectively. For the $\gamma$-ray bound the
differences are greater, ranging from 3.34-6.84$\sigma$ and
2.35-3.98$\sigma$ as described before.

In addition, we have done an analysis similar to the one before,
only that in this case we have calculated the differences in the
number of events considering the six different pairs of Doppler
calculation's types. This is displayed in Table~\ref{sigma2} and
Table~\ref{sigma3}, for a specific geometry, spherical and
cylindrical, respectively. There are only two favourable situations
for discriminating the Doppler factor models using the number of
events. This is for the case of a spherical geometry, the
$\gamma$-ray bound and without $\delta$ distribution, where we have
obtained 2.67$\sigma$ in separation for the events predicted using
$\delta_{EQ}$ and $\delta_{IC}$. The other case is when we compare
$\delta_{IC}$ and $\delta_{Var}$, where we have got 3.05$\sigma$.
For all the rest of pair comparisons we have got negligible
differences, less than 1.6$\sigma$ in the splitting of the number of
events.

In spite of not including variations in ``$a$'', the relation
between the radius and the length in the cylinder ($l=ra$), we have
checked that these changes will not lead to a relevant difference in
the number of events.

\begin{figure}[p]
\begin{center}
  \includegraphics[height=17.5cm,keepaspectratio=true]{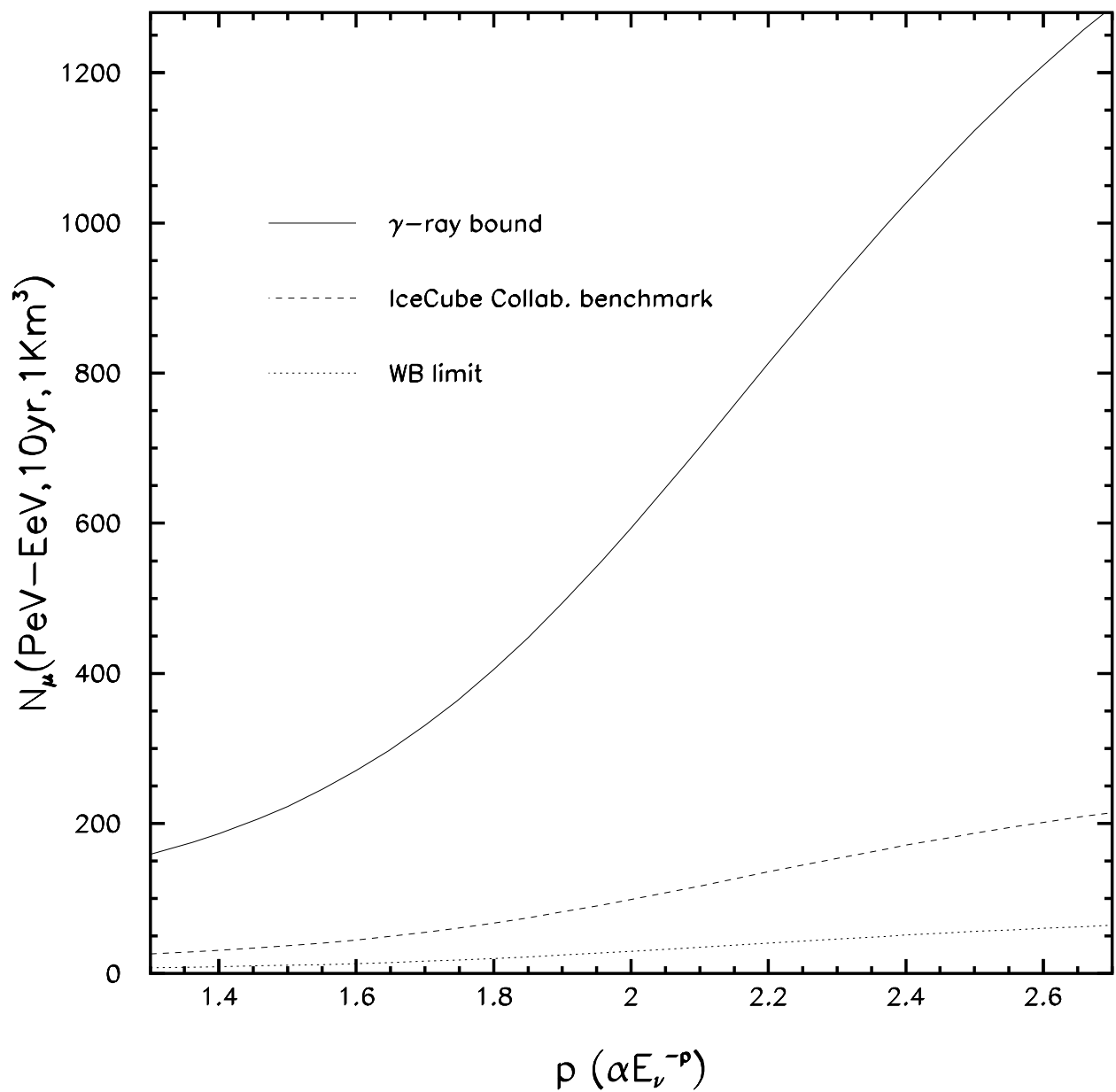}\\
  \caption{Expected total AGN muon neutrino induced events normalized with different bounds of the $\nu$ flux limit as 
a function of the exponent in the power law assumed in the AGN-$\numu$ flux.}
\label{spectrumvar}
\end{center}
\end{figure}

As complementary information, we show in Fig. \ref{spectrumvar} the
total number of expected events, for the three flux bounds, as a
function of the slope $p$ of the AGN
spectrum, ranging from 1.3 to 2.7. It is clear that in $p=2$ 
the number of events in this figure coincides
with the maximum number of events in our 
precedent figures. Thus, from this plot we can rescale 
the results given in our tables, only multiplying 
the $N_{\sigma}$ by $\sqrt{N_{\mu}(p)/N_{\mu}(p=2)}$, which at $p=2.7$ is 
around $\sqrt{2}$.

%%%%%%%%%%%%%%%%%%%%%%%%%%%%%%%%%%%%%%%%%%%%%%%%%%%%%%%%%%%%%%%%%%%%%%%
\section{\label{sec:con} Conclusions}
%%%%%%%%%%%%%%%%%%%%%%%%%%%%%%%%%%%%%%%%%%%%%%%%%%%%%%%%%%%%%%%%%%%%%%%

We have studied different neutrino production scenarios in AGNs for
the highest energy neutrinos (PeV-Eev) and their impact in the total
expected $\numu$-induced events in the IceCube detector, due to
variations in the emission region geometry, the Doppler factor
estimation model and the variability time. We have obtained the
total neutrino flux based on a sample of 39 AGN (i.e. point sources
observations), where the limits in the total neutrino flux were the
WB bound, the IceCube Collab. benchmark and the $\gamma$-ray
bound.

We have found as a global behaviour that if most AGN had long times
of burst, which seems natural to assume, we should be able to
distinguish, in some cases, among the different models. Otherwise,
for small times of burst, no information about these model
parameters could be extracted through neutrino observation in a
detector with the capabilities of IceCube.

More specifically, for a 10-day fluctuation time and the $\gamma$-ray
flux limit ($6\times 10^{-7}$ GeV$^{-1}$cm$^{-2}$s$^{-1}$sr$^{-1}$),
we could determine the emission region geometry after 10 years of
IceCube exposure. We observe a separation greater than 3$\sigma$
in the total number of events, between the cylindrical 
and spherical geometry, for all the $\delta$ cases. On the other hand,
if the flux bound was below the WB limit ($3\times 10^{-8}$
GeV$^{-1}$cm$^{-2}$s$^{-1}$sr$^{-1}$), the geometry would not
introduce major changes in the total number of events. When the 
benchmark is $1\times 10^{-7}$ GeV$^{-1}$cm$^{-2}$s$^{-1}$sr$^{-1}$, we 
are able to discriminate 
between geometries in half of the $\delta$ cases, without including 
the $\delta$ distribution, with a separation in the total number of 
events greater than 2$\sigma$.

It is important to note that it is possible to infere from
Fig.~\ref{eventvstimewb}, Fig.~\ref{eventvstimebe} and
Fig.~\ref{eventvstimegray}, any intermediate case (a mixture of both
geometries), between the cylindrical and spherical geometry for the
emission region. In this sense, how far or close we are from some of
these extreme behaviours (purely cylindrical or 
spherical) is a measure of the tendency of the sample.

For the Doppler factor estimation models, in most cases, the total
number of neutrino induced events gives no relevant information or,
in other words, the differences in $\delta$ do not affect this
number. Nonetheless, for a spherical emission region, $\Delta
t_{obs}=$10 days, the $\gamma$-ray bound and without considering 
the $\delta$ distribution, it is possible to distinguish 
between $\delta_{EQ}$-$\delta_{IC}$
and $\delta_{IC}$-$\delta_{Var}$ obtaining an event separation
greater than 2.5$\sigma$.

It is relevant to keep on mind that our analysis has considered that
the neutrino spectrum had slope $p=2$. Therefore, for a slope $p>2$,
the separation in terms of $N_\sigma$ must increase, which means 
that we are going to enlarge the number of cases where we can 
differentiate the geometries and the Doppler factor
estimations models.

As closing remarks, we must say 
that our work establish possible tendencies in the parameters in the AGNs. 
In fact, we expect that these
tendencies become more robust and measurable in the future, when we have a 
bigger sample containing the necessary information and accurate point 
sources observations.

%%%%%%%%%%%%%%%%%%%%%%%%%%%%%%%%%%%%%%%%%%%%%%%%%%%%%%%%%%%%%%%%%%%%%%%
%%%%%%%%%%%%%% Thanks
%%%%%%%%%%%%%%%%%%%%%%%%%%%%%%%%%%%%%%%%%%%%%%%%%%%%%%%%%%%%%%%%%%%%%%%

\acknowledgments

This work was supported by Direcci\'on Acad\'emica de
Investigaci\'on (\textbf{DAI}), Pontificia Universidad Cat\'olica
del Per\'u, 2004.

%%%%%%%%%%%%%%%%%%%%%%%%%%%%%%%%%%%%%%%%%%%%%%%%%%%%%%%%%%%%%%%%%%%%%%%
%%%%%%%%%%%%%% Beginning of References
%%%%%%%%%%%%%%%%%%%%%%%%%%%%%%%%%%%%%%%%%%%%%%%%%%%%%%%%%%%%%%%%%%%%%%%

%%%%%%%%%%%%%%%%%%%%%%%%%%%%%%%%%%%%%%%%%%%%%%%%%%%%%%%%%%%%%%%%%%%%%%%
%%%%%%%%%%%%%% End of References
%%%%%%%%%%%%%%%%%%%%%%%%%%%%%%%%%%%%%%%%%%%%%%%%%%%%%%%%%%%%%%%%%%%%%%%

\newpage

%%%%%%%%%%%%%%%%%%%%%%%%%%%%%%%%%%%%%%%%%%%%%%%%%%%%%%%%%%%%%%%%%%%%%
%  Appendix
%%%%%%%%%%%%%%%%%%%%%%%%%%%%%%%%%%%%%%%%%%%%%%%%%%%%%%%%%%%%%%%%%%%%%
\appendix

\section{\label{trans} Relativistic transformations}

For the energy and time transformation between rest-frame to
observer-frame, which includes not only the common relativistic
transformation but also the cosmological expansion, we use
\begin{equation}\label{ener} E'=\frac{(1+z)}{\delta}E_{obs}~, \end{equation}
\begin{equation}\label{time} \Delta t'=\frac{\delta}{(1+z)}\Delta t_{obs}~, \end{equation}
where $\delta$ is the relativistic Doppler factor, also known as
bulk Lorentz factor and z the AGN redshift \footnote{There should
not be any misleading impression with the fact that the redshift
means a receding object and most Doppler factors values an
approaching source, because the redshift is due to cosmological
effects and the Doppler factor to the moving knots in the jets.}.

There are luminosity transformations \cite{Urry} which consider the
observed integrated spectrum and solid angle distributions for
moving, isotropic sources and for a continuous standard conical
jets. However, in order to proceed in a more general way (i.e. model
independent) we consider the following jet photon luminosity
transformation
\begin{equation}\label{lum} L'_j=\left(\frac{1+z}{\delta}\right)^{2} L_{jobs}~.\end{equation}

%%%%%%%%%%%%%%%%%%%%%%%%%%%%%%%%%%%%%%%%%%%%%%%%%%%%%%%%%%%%%%%%%%%%%%%
\section{\label{sec:data} Outflow Angle, Doppler factors and Distributions}
%%%%%%%%%%%%%%%%%%%%%%%%%%%%%%%%%%%%%%%%%%%%%%%%%%%%%%%%%%%%%%%%%%%%%%%

In order to obtain a more accurate number in the exponent, $\gamma$,
in the power law formula ($\delta^{-\gamma}$) for each Doppler
Factor estimation model distribution, we enlarge our 39 data of the
Doppler factor estimation model (there is one for each source)
calculating the Doppler factor for each knot in the source. The
method for calculating these Doppler factors is described in the
following lines.

First, we calculate the apparent transverse velocity
$\beta_{app}(j,k)$ corresponding to each knot (index $k$) for its
AGN source (index $j$), given by~\footnote{The formula from
Pearson\&Zensus (1987) widely used in older papers is replaced by
Eqs. (\ref{app},\ref{dist}) which consider a non-zero cosmological
constant.}
\begin{equation}
\label{app} \beta_{app}(j,k)=\frac{\mu(j,k) d_{m}(j)}{c}~,
\end{equation}
where $\mu(j,k)$ is the proper motion for each knot in the source,
$d_{m}(j)$ is the ``proper motion distance'' to the source,
calculated using \cite{Carroll}
\begin{equation}\label{dist} d_{m}(j)=\frac{c}{H_{0}}\int^{Z(j)}_{0}\sqrt{(1+z)^{2}(1+\Omega_{m}z)-z(2+z)\Omega_{\Lambda}}dz ~,\end{equation}
with the Hubble constant $H_0$ = 71 Km $s^{-1}$ $Mpc^{-1}$, the
matter density $\Omega_{m}=0.27\pm0.04$, and dark energy density
$\Omega_{\Lambda}=0.73\pm0.04$, according to WMAP latest results
\cite{WMAP}.

Then, we calculate the angle to the line of sight,
$\theta(j,k)_{mod}$, for each proper motion component and its
corresponding Doppler factor for each estimation model (the index
$mod$ = EQ, IC, Var), using
\begin{equation}\label{theta}
\theta(j,k)_{mod}=\arctan\left[\frac{2\beta_{app}(j,k)}{\beta_{app}(j,k)^{2}+\delta(j)_{mod}^{2}-1}\right]~.
\end{equation}

In the case of Minimum Doppler factor, which
gives the minimum angle, we use
\begin{equation}\label{thatemin}\theta(j,k)_{min}=\operatorname{arccot} \left[\beta_{app}(j,k)\right]~. \end{equation}

>From the distribution of $\theta(j,k)_{mod}$ for each source, we
obtain its mean value $\left<\theta\right>_{(j)mod}$ and assign it
to the source, which is well justified \cite{K98}. Thus, we can
construct the Doppler factor distribution for each model in terms of
the Doppler factor of the knots for each AGN source.

\begin{equation}\label{delta} \delta(j,k)_{mod}=\frac{\sqrt{1-\beta(j,k)_{mod}^{2}}}{1-\beta(j,k)_{mod}\cos \left<\theta\right>_{(j)mod}}~,
\end{equation}
where
\begin{equation}\label{beta} \beta(j,k)_{mod}=\frac{\beta_{app}(j,k)}{\beta_{app}(j,k)\cos \left<\theta\right>_{(j)mod}+\sin \left<\theta\right>_{(j)mod}}~. \end{equation}

We discard Doppler factors that could correspond to jets affected by
significant bends, twists and inward moving features.

\newpage
%%%%%%%%%%%%%%%%%%%%%%%%%%%%%%%%%%%%%%%%%%%%%%%%%%%%%%%%%%%%%%%%%%%%%
\section{The AGN Sample:~Model Parameters}
%%%%%%%%%%%%%%%%%%%%%%%%%%%%%%%%%%%%%%%%%%%%%%%%%%%%%%%%%%%%%%%%%%%%%

\renewcommand{\arraystretch}{.5}
\setlength\tabcolsep{5pt}

\begin{center}
\begin{longtable}[H]{ >{\scriptsize}c >{\scriptsize}c >{\scriptsize}c >{\scriptsize}c
>{\scriptsize}c >{\scriptsize}c >{\scriptsize}c >{\scriptsize}c
>{\scriptsize}c >{\scriptsize}c}
 \caption{BL Lacs}\label{tableblazar}\\
 \hline \hline Source&Alias&Redshift&Lum.&Comp.&$\mu_r$&$\delta_{EQ}$&$\delta_{IC}$&$\delta_{Var}$&Ref.\\ &  &  & (W $Hz^{-1}$)  &  & (mas
 $yr^{-1}$)&\\
 (1) & (2) & (3) & (4) & (5) & (6) & (7) & (8) & (9) & (10)
 \endfirsthead
\caption{-- Continued}\\
 \hline \hline Source&Alias&Redshift&Lum.&Comp.&$\mu_r$&$\delta_{EQ}$&$\delta_{IC}$&$\delta_{Var}$&Ref.\\ &  &  & (W $Hz^{-1}$)  &  & (mas $yr^{-1}$) &
 \\ (1) & (2) & (3) & (4) & (5) & (6) & (7) & (8) & (9) & (10)\\\hline\endhead \hline

0235+164    &   OD 160  &   0.94    &   1.2 x   $   10^{    27  }   $   &   . . .   &   2.35    &   3.7 &   5   &   16.32   &   G93 \\
    &       &       &                               &   B1  &   0.93    &       &       &       &   J01 \\
    &       &       &                               &   B2  &   0.61    &       &       &       &   J01 \\
0735+178    &   OI 158  &   0.424   &   2.6 x   $   10^{    26  }   $   &   NE  &   0.18    &   6.6 &   5.6 &   3.17    &   C88 \\
    &       &       &                               &   . . .   &   0.64    &       &       &       &   G93 \\
    &       &       &                               &   K1/U1   &   0.14    &       &       &       &   H00 \\
    &       &       &                               &   B   &   0.17    &       &       &       &   K04 \\
    &       &       &                               &   C   &   0.64    &       &       &       &   K04 \\
    &       &       &                               &   D   &   -0.18   &       &       &       &   K04 \\
    &       &       &                               &   C0  &   0.44    &       &       &       &   V94 \\
0754+100    &       &   0.266   &   1.3 x   $   10^{    27  }   $   &   B   &   0.70    &   0.48    &   0.85    &   5.52    &   K04 \\
    &       &       &                               &   C   &   0.05    &       &       &       &   K04 \\
0851+202    &   OJ 287  &   0.306   &   2.4 x   $   10^{    26  }   $   &   SW1-2   &   0.28    &   12  &   6.8 &   18.03   &   C88 \\
    &       &       &                               &   . . .   &   0.37    &       &       &       &   G93 \\
    &       &       &                               &   . . .   &   0.23    &       &       &       &   G97 \\
    &       &       &                               &   K3/U3   &   1.01    &       &       &       &   H00 \\
    &       &       &                               &   B3  &   0.43    &       &       &       &   J01 \\
    &       &       &                               &   B2  &   0.54    &       &       &       &   J01 \\
    &       &       &                               &   B1+D1+F1    &   0.67    &       &       &       &   J01 \\
    &       &       &                               &   C   &   0.52    &       &       &       &   K04 \\
    &       &       &                               &   D   &   0.37    &       &       &       &   K04 \\
    &       &       &                               &   E   &   0.31    &       &       &       &   K04 \\
    &       &       &                               &   K1  &   0.20    &       &       &       &   V94 \\
    &       &       &                               &   K2  &   0.27    &       &       &       &   V94 \\
1219+285    &   W Comae &   0.102   &   1.1 x   $   10^{    25  }   $   &   B9  &   0.13    &   0.14    &   0.15    &   1.56    &   J01 \\
    &       &       &                               &   B3+B8   &   0.32    &       &       &       &   J01 \\
    &       &       &                               &   B7  &   0.60    &       &       &       &   J01 \\
    &       &       &                               &   B2+B6   &   0.47    &       &       &       &   J01 \\
    &       &       &                               &   B1+B5   &   0.50    &       &       &       &   J01 \\
    &       &       &                               &   B   &   0.08    &       &       &       &   K04 \\
    &       &       &                               &   C   &   0.02    &       &       &       &   K04 \\
    &       &       &                               &   D   &   0.48    &       &       &       &   K04 \\
1749+096    &   4C 73.18    &   0.32    &   1.3 x   $   10^{    27  }   $   &   K3/U3   &   0.06    &   18  &   11  &   15.85   &   H00 \\
    &       &       &                               &   K2/U2   &   0.22    &       &       &       &   H00 \\
    &       &       &                               &   K1/U1   &   0.25    &       &       &       &   H00 \\
    &       &       &                               &   B   &   0.06    &       &       &       &   K04 \\
    &       &       &                               &   C   &   0.15    &       &       &       &   K04 \\
1803+784    &       &   0.68    &   1.8 x   $   10^{    27  }   $   &   . . .   &   0.25    &   5.6 &   6.6 &   6.45    &   G93 \\
    &       &       &                               &   . . .   &   0.00    &       &       &       &   G97 \\
    &       &       &                               &   B   &   -0.01   &       &       &       &   K04 \\
    &       &       &                               &   . . .   &   0.00    &       &       &       &   V94 \\
1807+698    &   3C 371  &   0.05    &   7.3 x   $   10^{    24  }   $   &   . . .   &   6.18    &   0.44    &   0.54    &   1.8 &   G93 \\
    &       &       &                               &   B   &   0.01    &       &       &       &   K04 \\
    &       &       &                               &   C   &   0.12    &       &       &       &   K04 \\
    &       &       &                               &   D   &   0.85    &       &       &       &   K04 \\
    &       &       &                               &   E   &   0.26    &       &       &       &   K04 \\
2007+777    &       &   0.342   &   2.7 x   $   10^{    26  }   $   &   . . .   &   0.25    &   2.9 &   3.6 &   5.13    &   G93 \\
    &       &       &                               &   . . .   &   0.18    &       &       &       &   G97 \\
    &       &       &                               &   U3  &   0.04    &       &       &       &   H00 \\
    &       &       &                               &   U2  &   0.02    &       &       &       &   H00 \\
    &       &       &                               &   U1  &   0.20    &       &       &       &   H00 \\
    &       &       &                               &   D   &   -0.04   &       &       &       &   K04 \\
    &       &       &                               &   C2  &   0.18    &       &       &       &   V94 \\
2200+420    &   BL Lac  &   0.069   &   3.7 x   $   10^{    25  }   $   &   1-4 &   0.76    &   5.2 &   3.4 &   3.91    &   C88 \\
    &       &       &                               &   . . .   &   1.62    &       &       &       &   G93 \\
    &       &       &                               &   . . .   &   0.99    &       &       &       &   G97 \\
    &       &       &                               &   S10 &   1.15    &       &       &       &   J01 \\
    &       &       &                               &   S9  &   1.91    &       &       &       &   J01 \\
    &       &       &                               &   S8  &   0.71    &       &       &       &   J01 \\
    &       &       &                               &   S7  &   1.36    &       &       &       &   J01 \\
    &       &       &                               &   B   &   1.41    &       &       &       &   K04 \\
    &       &       &                               &   C   &   1.12    &       &       &       &   K04 \\
    &       &       &                               &   D   &   0.99    &       &       &       &   K04 \\
    &       &       &                               &   E   &   1.09    &       &       &       &   K04 \\
    &       &       &                               &   S1  &   1.20    &       &       &       &   V94 \\
    &       &       &                               &   S2  &   1.10    &       &       &       &   V94 \\
    &       &       &                               &   S3  &   1.10    &       &       &       &   V94 \\
    &       &       &                               &   S5  &   1.00    &       &       &       &   V94 \\

\\ \hline
\end{longtable}
\end{center}

\begin{center}
\begin{longtable}[H]{ >{\scriptsize}c >{\scriptsize}c >{\scriptsize}c >{\scriptsize}c >{\scriptsize}c >{\scriptsize}c >{\scriptsize}c >{\scriptsize}c >{\scriptsize}c >{\scriptsize}c}
 \caption{Quasars}\label{tablequasars}\\
 \hline \hline Source&Alias&Redshift&Lum.&Comp.&$\mu_r$&$\delta_{EQ}$&$\delta_{IC}$&$\delta_{Var}$&Ref.\\ &  &  & (W $Hz^{-1}$)  &  & (mas
 $yr^{-1}$)&\\
 (1) & (2) & (3) & (4) & (5) & (6) & (7) & (8) & (9) & (10)
 \endfirsthead
\caption{-- Continued}\\
 \hline \hline Source&Alias&Redshift&Lum.&Comp.&$\mu_r$&$\delta_{EQ}$&$\delta_{IC}$&$\delta_{Var}$&Ref.\\ &  &  & (W $Hz^{-1}$)  &  & (mas $yr^{-1}$) &
 \\ (1) & (2) & (3) & (4) & (5) & (6) & (7) & (8) & (9) & (10)\\\hline\endhead \hline

0016+731    &       &   1.781   &   3.2 x   $   10^{    27  }   $   &   . . .   &   0.31    &   5.7 &   7.9 &   18.37   &   G93 \\
    &       &       &                               &   . . .   &   0.30    &       &       &       &   G97 \\
    &       &       &                               &   B   &   0.07    &       &       &       &   K04 \\
    &       &       &                               &   . . .   &   0.22    &       &       &       &   V94 \\
0106+013    &   4C 01.02    &   2.107   &   7.9 x   $   10^{    27  }   $   &   . . .   &   0.28    &   14  &   15  &   8.62    &   G93 \\
    &       &       &                               &   B   &   0.28    &       &       &       &   K04 \\
    &       &       &                               &   C   &   0.27    &       &       &       &   K04 \\
    &       &       &                               &   C2  &   0.20    &       &       &       &   V94 \\
0133+476    &   DA 55   &   0.859   &   3   x   $   10^{    27  }   $   &   B   &   0.04    &   12  &   13  &   7.09    &   K04 \\
0212+735    &       &   2.367   &   1.9 x   $   10^{    28  }   $   &   . . .   &   0.09    &   5.6 &   7.1 &   4.16    &   C88 \\
    &       &       &                               &   . . .   &   0.13    &       &       &       &   G93 \\
    &       &       &                               &   . . .   &   0.13    &       &       &       &   G97 \\
    &       &       &                               &   B   &   0.08    &       &       &       &   K04 \\
    &       &       &                               &   C   &   -0.01   &       &       &       &   K04 \\
    &       &       &                               &   C2  &   0.09    &       &       &       &   V94 \\
    &       &       &                               &   C3  &   0.07    &       &       &       &   V94 \\
    &       &       &                               &   C4  &   -0.14   &       &       &       &   V94 \\
0234+285    &   CTD 20  &   1.213   &   3.8 x   $   10^{    27  }   $   &   . . .   &   0.36    &   7.1 &   13  &   7.29    &   G97 \\
    &       &       &                               &   B   &   0.23    &       &       &       &   K04 \\
    &       &       &                               &   . . .   &   0.30    &       &       &       &   V94 \\
0333+321    &   NRAO 140    &   1.263   &   3.2 x   $   10^{    27  }   $   &   B   &   0.15    &   29  &   13  &   6.48    &   C88 \\
    &       &       &                               &   . . .   &   0.21    &       &       &       &   G93 \\
    &       &       &                               &   . . .   &   0.18    &       &       &       &   G97 \\
    &       &       &                               &   B   &   0.18    &       &       &       &   K04 \\
    &       &       &                               &   C   &   0.20    &       &       &       &   K04 \\
    &       &       &                               &   D   &   0.40    &       &       &       &   K04 \\
    &       &       &                               &   E   &   -0.08   &       &       &       &   K04 \\
    &       &       &                               &   B   &   0.15    &       &       &       &   V94 \\
0336-019    &   CTA 26  &   0.852   &   3   x   $   10^{    27  }   $   &   C4  &   0.18    &   13  &   12  &   19.01   &   J01 \\
    &       &       &                               &   C3  &   0.42    &       &       &       &   J01 \\
    &       &       &                               &   B   &   0.22    &       &       &       &   K04 \\
    &       &       &                               &   C   &   0.21    &       &       &       &   K04 \\
0420-014    &   OA 129  &   0.915   &   6.3 x   $   10^{    27  }   $   &   B   &   0.20    &   21  &   13  &   11.72   &   J01 \\
    &       &       &                               &   B   &   0.03    &       &       &       &   K04 \\
    &       &       &                               &   C   &   0.29    &       &       &       &   K04 \\
0528+134    &   PKS 0528+134    &   2.07    &   4.6 x   $   10^{    28  }   $   &   K2/U2   &   0.16    &   0.71    &   2   &   14.22   &   H00 \\
    &       &       &                               &   F2  &   0.40    &       &       &       &   J01 \\
    &       &       &                               &   B1+F3   &   0.26    &       &       &       &   J01 \\
    &       &       &                               &   B2+F4   &   0.33    &       &       &       &   J01 \\
    &       &       &                               &   B3  &   0.19    &       &       &       &   J01 \\
    &       &       &                               &   B4  &   0.15    &       &       &       &   J01 \\
    &       &       &                               &   B   &   0.08    &       &       &       &   K04 \\
0552+398    &   DA 193  &   2.365   &   3.6 x   $   10^{    28  }   $   &   . . .   &   0.06    &   1.2 &   2.2 &   14.2    &   G93 \\
    &       &       &                               &   . . .   &   0.06    &       &       &       &   G97 \\
    &       &       &                               &   . . .   &   0.04    &       &       &       &   V94 \\
0804+499    &       &   1.432   &   1.9 x   $   10^{    27  }   $   &   B   &   0.13    &   21  &   16  &   26.21   &   K04 \\
0836+710    &   4C 71.07    &   2.17    &   1.39    x   $   10^{    28  }   $   &   . . .   &   0.35    &   8.5 &   6.7 &   10.67   &   G93 \\
    &       &       &                               &   . . .   &   0.27    &       &       &       &   G97 \\
    &       &       &                               &   B   &   0.24    &       &       &       &   J01 \\
    &       &       &                               &   B   &   0.23    &       &       &       &   V94 \\
    &       &       &                               &   D   &   0.14    &       &       &       &   V94 \\
0923+392    &   4C 39.28    &   0.698   &   1   x   $   10^{    28  }   $   &   A-C &   0.01    &   6.9 &   8.9 &   2.25    &   C88 \\
    &       &       &                               &   B   &   0.16    &       &       &       &   C88 \\
    &       &       &                               &   . . .   &   0.22    &       &       &       &   G93 \\
    &       &       &                               &   . . .   &   0.19    &       &       &       &   G97 \\
    &       &       &                               &   B   &   0.07    &       &       &       &   K04 \\
    &       &       &                               &   C   &   0.01    &       &       &       &   K04 \\
    &       &       &                               &   B   &   0.18    &       &       &       &   V94 \\
    &       &       &                               &   D   &   0.02    &       &       &       &   V94 \\
1156+295    &   4C 29.45    &   0.729   &   1   x   $   10^{    27  }   $   &   . . .   &   1.58    &   2.2 &   4.9 &   9.42    &   G93 \\
    &       &       &                               &   . . .   &   1.24    &       &       &       &   G97 \\
    &       &       &                               &   B2  &   0.34    &       &       &       &   J01 \\
    &       &       &                               &   B3  &   0.54    &       &       &       &   J01 \\
    &       &       &                               &   B   &   0.22    &       &       &       &   K04 \\
    &       &       &                               &   . . .   &   1.15    &       &       &       &   V94 \\
1226+023    &   3C 273  &   0.158   &   1.5 x   $   10^{    27  }   $   &   C3  &   0.79    &   8.3 &   4.6 &   5.71    &   C88 \\
    &       &       &                               &   C4  &   0.99    &       &       &       &   C88 \\
    &       &       &                               &   C5  &   1.20    &       &       &       &   C88 \\
    &       &       &                               &   C7a &   0.76    &       &       &       &   C88 \\
    &       &       &                               &   . . .   &   1.62    &       &       &       &   G93 \\
    &       &       &                               &   . . .   &   0.88    &       &       &       &   G97 \\
    &       &       &                               &   K10/U10 &   0.77    &       &       &       &   H00 \\
    &       &       &                               &   K9/U9   &   0.94    &       &       &       &   H00 \\
    &       &       &                               &   K8/U8   &   1.15    &       &       &       &   H00 \\
    &       &       &                               &   K7/U7   &   1.06    &       &       &       &   H00 \\
    &       &       &                               &   K4/U4   &   0.99    &       &       &       &   H00 \\
    &       &       &                               &   B5+E2   &   0.33    &       &       &       &   J01 \\
    &       &       &                               &   B4+E1   &   0.85    &       &       &       &   J01 \\
    &       &       &                               &   B3  &   0.66    &       &       &       &   J01 \\
    &       &       &                               &   B2  &   0.61    &       &       &       &   J01 \\
    &       &       &                               &   B1+G1   &   0.68    &       &       &       &   J01 \\
    &       &       &                               &   H5+K6+D1    &   1.60    &       &       &       &   J01 \\
    &       &       &                               &   H3+K5+D2    &   0.70    &       &       &       &   J01 \\
    &       &       &                               &   B   &   1.05    &       &       &       &   K04 \\
    &       &       &                               &   C   &   1.36    &       &       &       &   K04 \\
    &       &       &                               &   D   &   0.83    &       &       &       &   K04 \\
    &       &       &                               &   E   &   1.27    &       &       &       &   K04 \\
    &       &       &                               &   F   &   0.79    &       &       &       &   K04 \\
    &       &       &                               &   G   &   0.41    &       &       &       &   K04 \\
    &       &       &                               &   I   &   0.88    &       &       &       &   K04 \\
    &       &       &                               &   C2  &   1.15    &       &       &       &   V94 \\
    &       &       &                               &   C3  &   0.79    &       &       &       &   V94 \\
    &       &       &                               &   C4  &   0.99    &       &       &       &   V94 \\
    &       &       &                               &   C5  &   1.20    &       &       &       &   V94 \\
    &       &       &                               &   C7  &   0.65    &       &       &       &   V94 \\
    &       &       &                               &   C7a &   0.76    &       &       &       &   V94 \\
    &       &       &                               &   C8  &   0.92    &       &       &       &   V94 \\
    &       &       &                               &   C9  &   0.82    &       &       &       &   V94 \\
1253-055    &   3C 279  &   0.538   &   1.3 x   $   10^{    28  }   $   &   . . .   &   0.50    &   14  &   14  &   16.77   &   C88 \\
    &       &       &                               &   B2  &   0.11    &       &       &       &   C88 \\
    &       &       &                               &   . . .   &   0.68    &       &       &       &   G93 \\
    &       &       &                               &   . . .   &   0.14    &       &       &       &   G97 \\
    &       &       &                               &   K4/U4   &   0.17    &       &       &       &   H00 \\
    &       &       &                               &   K1/U1   &   0.25    &       &       &       &   H00 \\
    &       &       &                               &   B3  &   0.18    &       &       &       &   J01 \\
    &       &       &                               &   B2  &   0.17    &       &       &       &   J01 \\
    &       &       &                               &   E2+B1   &   0.27    &       &       &       &   J01 \\
    &       &       &                               &   D   &   0.31    &       &       &       &   J01 \\
    &       &       &                               &   B   &   0.28    &       &       &       &   K04 \\
    &       &       &                               &   C3  &   0.12    &       &       &       &   V94 \\
    &       &       &                               &   . . .   &   0.50    &       &       &       &   V94 \\
1308+326    &   OR 017  &   0.997   &   5.8 x   $   10^{    27  }   $   &   K2/U2   &   0.19    &   4.3 &   5.2 &   11.38   &   H00 \\
    &       &       &                               &   K1/U1   &   0.96    &       &       &       &   H00 \\
    &       &       &                               &   . . .   &   0.45    &       &       &       &   G97 \\
    &       &       &                               &   K1  &   0.13    &       &       &       &   V94 \\
    &       &       &                               &   K2  &   0.75    &       &       &       &   V94 \\
    &       &       &                               &   K3  &   0.29    &       &       &       &   V94 \\
    &       &       &                               &   B   &   0.31    &       &       &       &   K04 \\
1510-089    &   4C 09.56    &   0.36    &   3.4 x   $   10^{    26  }   $   &   K3/U3   &   0.45    &   10  &   11  &   13.18   &   H00 \\
    &       &       &                               &   B1  &   0.51    &       &       &       &   J01 \\
    &       &       &                               &   D1  &   0.28    &       &       &       &   J01 \\
    &       &       &                               &   D2  &   0.63    &       &       &       &   J01 \\
    &       &       &                               &   B   &   0.85    &       &       &       &   K04 \\
    &       &       &                               &   C   &   0.57    &       &       &       &   K04 \\
1633+382    &   4C 38.41    &   1.807   &   6.5 x   $   10^{    27  }   $   &   B3  &   0.14    &   0.83    &   2.2 &   8.83    &   J01 \\
    &       &       &                               &   B1  &   0.20    &       &       &       &   J01 \\
    &       &       &                               &   B   &   0.15    &       &       &       &   K04 \\
    &       &       &                               &   C   &   0.10    &       &       &       &   K04 \\
1641+399    &   3C 345  &   0.594   &   5.9 x   $   10^{    27  }   $   &   C2  &   0.48    &   1.5 &   4.1 &   7.45    &   C88 \\
    &       &       &                               &   C3  &   0.30    &       &       &       &   C88 \\
    &       &       &                               &   C4  &   0.30    &       &       &       &   C88 \\
    &       &       &                               &   . . .   &   0.66    &       &       &       &   G93 \\
    &       &       &                               &   . . .   &   0.33    &       &       &       &   G97 \\
    &       &       &                               &   B   &   0.49    &       &       &       &   K04 \\
    &       &       &                               &   C   &   0.37    &       &       &       &   K04 \\
    &       &       &                               &   C2  &   0.47    &       &       &       &   V94 \\
    &       &       &                               &   C3  &   0.30    &       &       &       &   V94 \\
    &       &       &                               &   C4  &   0.31    &       &       &       &   V94 \\
    &       &       &                               &   C5  &   0.23    &       &       &       &   V94 \\
1928+738    &   4C 73.18    &   0.303   &   6.2 x   $   10^{    26  }   $   &   A1-4    &   0.60    &   3.4 &   3.4 &   3.71    &   C88 \\
    &       &       &                               &   . . .   &   0.81    &       &       &       &   G93 \\
    &       &       &                               &   . . .   &   0.39    &       &       &       &   G97 \\
    &       &       &                               &   B   &   0.29    &       &       &       &   K04 \\
    &       &       &                               &   C   &   0.30    &       &       &       &   K04 \\
    &       &       &                               &   E   &   0.12    &       &       &       &   K04 \\
    &       &       &                               &   A1  &   0.32    &       &       &       &   V94 \\
    &       &       &                               &   B   &   0.37    &       &       &       &   V94 \\
    &       &       &                               &   C   &   0.34    &       &       &       &   V94 \\
    &       &       &                               &   C2  &   0.51    &       &       &       &   V94 \\
    &       &       &                               &   C3  &   0.57    &       &       &       &   V94 \\
    &       &       &                               &   C4  &   0.40    &       &       &       &   V94 \\
    &       &       &                               &   C6  &   0.40    &       &       &       &   V94 \\
    &       &       &                               &   C7  &   0.60    &       &       &       &   V94 \\
    &       &       &                               &   C9  &   0.31    &       &       &       &   V94 \\
2021+614    &   OW 637  &   0.227   &   2.6 x   $   10^{    26  }   $   &   . . .   &   0.04    &   0.9 &   1.1 &   1.59    &   C88 \\
    &       &       &                               &   . . .   &   0.05    &       &       &       &   G93 \\
    &       &       &                               &   . . .   &   0.02    &       &       &       &   G97 \\
    &       &       &                               &   B   &   0.04    &       &       &       &   K04 \\
    &       &       &                               &   C   &   0.05    &       &       &       &   K04 \\
    &       &       &                               &   . . .   &   0.02    &       &       &       &   V94 \\
2134+004    &   PHL 61  &   1.932   &   2.9 x   $   10^{    28  }   $   &   . . .   &   0.01    &   15  &   27  &   11.49   &   C88 \\
    &       &       &                               &   . . .   &   0.01    &       &       &       &   G93 \\
    &       &       &                               &   B   &   0.02    &       &       &       &   K04 \\
2145+067    &   4C 06.69    &   0.999   &   1.1 x   $   10^{    28  }   $   &   B   &   -0.01   &   15  &   21  &   7.81    &   K04 \\
    &       &       &                               &   C   &   0.03    &       &       &       &   K04 \\
    &       &       &                               &   D   &   0.03    &       &       &       &   K04 \\
2223-052    &   3C 446  &   1.404   &   1.2 x   $   10^{    28  }   $   &   . . .   &   0.14    &   18  &   16  &   11.39   &   G93 \\
    &       &       &                               &   B   &   0.49    &       &       &       &   K04 \\
    &       &       &                               &   C   &   0.31    &       &       &       &   K04 \\
2230+114    &   CTA 102 &   1.037   &   4.3 x   $   10^{    27  }   $   &   . . .   &   0.65    &   1   &   1.5 &   14.23   &   C88 \\
    &       &       &                               &   . . .   &   0.69    &       &       &       &   G93 \\
    &       &       &                               &   B3  &   0.25    &       &       &       &   J01 \\
    &       &       &                               &   B2  &   0.34    &       &       &       &   J01 \\
    &       &       &                               &   B1  &   0.33    &       &       &       &   J01 \\
    &       &       &                               &   B   &   0.03    &       &       &       &   K04 \\
    &       &       &                               &   C   &   -0.05   &       &       &       &   K04 \\
    &       &       &                               &   E   &   -0.23   &       &       &       &   K04 \\
2251+158    &   3C 454.3    &   0.859   &   1.2 x   $   10^{    28  }   $   &   2   &   0.05    &   5.6 &   4.6 &   21.84   &   C88 \\
    &       &       &                               &   4   &   0.35    &       &       &       &   C88 \\
    &       &       &                               &   . . .   &   0.48    &       &       &       &   G93 \\
    &       &       &                               &   . . .   &   0.09    &       &       &       &   G97 \\
    &       &       &                               &   B3  &   0.14    &       &       &       &   J01 \\
    &       &       &                               &   B2  &   0.34    &       &       &       &   J01 \\
    &       &       &                               &   B1  &   0.53    &       &       &       &   J01 \\
    &       &       &                               &   B   &   0.04    &       &       &       &   K04 \\
    &       &       &                               &   2   &   0.05    &       &       &       &   V94 \\
    &       &       &                               &   3   &   -0.05   &       &       &       &   V94 \\
    &       &       &                               &   4   &   0.35    &       &       &       &   V94 \\
    &       &       &                               &   5   &   0.21    &       &       &       &   V94 \\

\\ \hline
\end{longtable}
\end{center}

\begin{center}
\begin{longtable}[H]{ >{\scriptsize}c >{\scriptsize}c >{\scriptsize}c >{\scriptsize}c >{\scriptsize}c >{\scriptsize}c >{\scriptsize}c >{\scriptsize}c >{\scriptsize}c >{\scriptsize}c}
 \caption{Radio Galaxy \& Seyfert}\label{tableradiogalaxies}\\
 \hline \hline Source&Alias&Redshift&Lum.&Comp.&$\mu_r$&$\delta_{EQ}$&$\delta_{IC}$&$\delta_{Var}$&Ref.\\ &  &  & (W $Hz^{-1}$)  &  & (mas
 $yr^{-1}$)&\\
 (1) & (2) & (3) & (4) & (5) & (6) & (7) & (8) & (9) & (10)
 \endfirsthead
\caption{-- Continued}\\
 \hline \hline Source&Alias&Redshift&Lum.&Comp.&$\mu_r$&$\delta_{EQ}$&$\delta_{IC}$&$\delta_{Var}$&Ref.\\ &  &  & (W $Hz^{-1}$)  &  & (mas $yr^{-1}$) &
 \\ (1) & (2) & (3) & (4) & (5) & (6) & (7) & (8) & (9) & (10)\\\hline\endhead \hline

1845+797    &   3C 390.3    &   0.057   &   2.9 x   $   10^{    24  }   $   &   . . .   &   0.98    &   0.37    &   0.38    &   1.16    &   G93 \\
    &       &       &                               &   B   &   0.54    &       &       &       &   K04 \\
    &       &       &                               &   C   &   0.60    &       &       &       &   K04 \\
0430+052    &   3C 120  &   0.033   &   8.2 x   $   10^{    24  }   $   &   A   &   1.35    &   11  &   4.1 &   0.98    &   C88 \\
    &       &       &                               &   B   &   2.53    &       &       &       &   C88 \\
    &       &       &                               &   C   &   2.47    &       &       &       &   C88 \\
    &       &       &                               &   D   &   2.66    &       &       &       &   C88 \\
    &       &       &                               &   E   &   2.54    &       &       &       &   C88 \\
    &       &       &                               &   . . .   &   3.55    &       &       &       &   G93 \\
    &       &       &                               &   . . .   &   2.06    &       &       &       &   G97 \\
    &       &       &                               &   K1B/U1B &   1.62    &       &       &       &   H00 \\
    &       &       &                               &   K1A/U1A &   2.22    &       &       &       &   H00 \\
    &       &       &                               &   B   &   1.77    &       &       &       &   K04 \\
    &       &       &                               &   C   &   1.80    &       &       &       &   K04 \\
    &       &       &                               &   D   &   1.36    &       &       &       &   K04 \\
    &       &       &                               &   G   &   1.59    &       &       &       &   K04 \\
    &       &       &                               &   I   &   1.51    &       &       &       &   K04 \\
    &       &       &                               &   H   &   2.08    &       &       &       &   K04 \\
    &       &       &                               &   A   &   1.35    &       &       &       &   V94 \\
    &       &       &                               &   B   &   2.53    &       &       &       &   V94 \\
    &       &       &                               &   C   &   2.47    &       &       &       &   V94 \\
    &       &       &                               &   D   &   2.66    &       &       &       &   V94 \\
    &       &       &                               &   E   &   2.54    &       &       &       &   V94 \\

\\ \hline
\end{longtable}
\end{center}

\footnotesize Note.- The acronyms used in the references for the
proper motion, given in column (10), correspond to: C88
(\cite{C88}), G93 (\cite{G93}), G97 (\cite{G97}), H00 (\cite{H00}),
J01 (\cite{J01}), K04 (\cite{K04}), and V94 (\cite{V94}).

%%%%%%%%%%%%%%%%%%%%%%%%%%%%%%%%%%%%%%%%%%%%%%%%%%%%%%%%%%%%%%%%%%%%%%%
%%%%%%%%%%%%%% End of Appendix
%%%%%%%%%%%%%%%%%%%%%%%%%%%%%%%%%%%%%%%%%%%%%%%%%%%%%%%%%%%%%%%%%%%%%%%
\newpage

\begin{figure}[h]
\begin{center}
  \includegraphics[width=9cm,keepaspectratio=true]{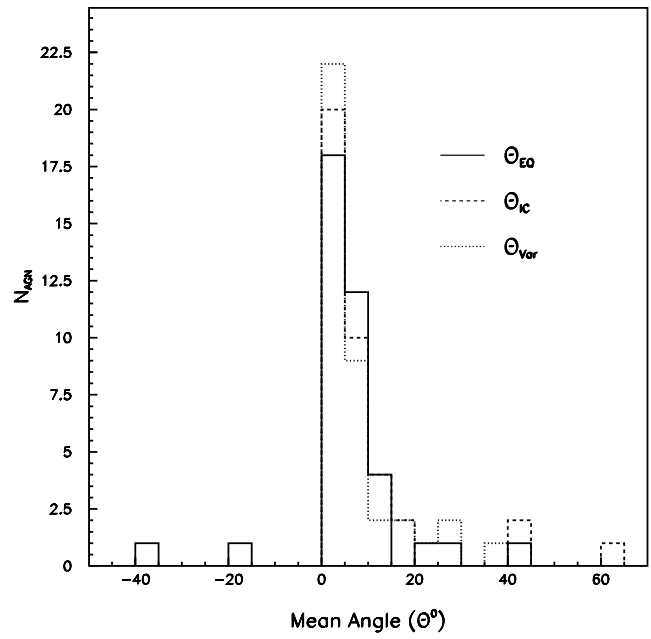}\\
  \caption{Distribution of the mean angle to the line of sight in the sample.}\label{histangle}
\end{center}
\end{figure}

\newpage

\newpage

\newpage

\newpage

\begin{table}[p]
\begin{center} {\footnotesize
\begin{tabular*}{0.75\textwidth}{@{\extracolsep{\fill}} |c!{\vrule width 1.5pt}cccc|cccc|}
\hline
\multirow{2}{2.3cm}{~~$\nu$ flux limit}& \multicolumn{4}{c|}{Without $\delta$ distribution} & \multicolumn{4}{c|}{Using $\delta$ distribution} \\
   &\multicolumn{1}{c}{$\delta_{EQ}$} &
\multicolumn{1}{c}{$\delta_{IC}$} &
\multicolumn{1}{c}{$\delta_{Var}$} &
\multicolumn{1}{c|}{$\delta_{Min}$}&
\multicolumn{1}{c}{$\delta_{EQ}$} &
\multicolumn{1}{c}{$\delta_{IC}$} &
\multicolumn{1}{c}{$\delta_{Var}$} &
\multicolumn{1}{c|}{$\delta_{Min}$}\\
\hline \hline$WB$ & 1.53 & 0.75 & 1.45 & 1.08 & 0.81 & 0.56 & 0.89 & 0.53\\
\hline $IceCube$ & 2.79 & 1.36 & 2.64 & 1.97 & 1.48 & 1.02 & 1.62 & 0.96\\
\hline $\gamma-ray$ & 6.84 & 3.34 & 6.47 & 4.82 & 3.62 & 2.51 & 3.98 & 2.35\\
\hline
\end{tabular*} }
\end{center}
\caption{\footnotesize Separation in terms of $N_\sigma$ between 
the number of events 
obtained for the spherical and
cylindrical geometries, using the three $\nu$ flux limits, all $\delta$ cases and $\Delta t_{obs}=$ 10days.} \label{sigma}
\end{table}

\begin{table}[p]
\begin{center} {\footnotesize
\begin{tabular*}{0.75\textwidth}{@{\extracolsep{\fill}} |c!{\vrule width 1.5pt}cccccc|}
\hline
\multirow{2}{2.3cm}{~~$\nu$ flux limit}& \multicolumn{6}{c|}{Without $\delta$ distribution} \\
   &\multicolumn{1}{c}{EQ-IC} &
\multicolumn{1}{c}{EQ-Var} & \multicolumn{1}{c}{EQ-Min} &
\multicolumn{1}{c}{IC-Var}& \multicolumn{1}{c}{IC-Min} &
\multicolumn{1}{c|}{Var-Min} \\
\hline \hline$WB$ & 0.68 & 0.04 & 0.35 & 0.59 & 0.31 & 0.31  \\
\hline $IceCube$ & 1.25& 0.08&0.64 &1.09 &0.56  &0.56  \\
\hline $\gamma-ray$ & 3.05& 0.18& 1.57&2.67 &1.38 &1.38  \\
\hline\hline
& \multicolumn{6}{c|}{Using $\delta$ distribution} \\
   \hline \hline$WB$ &  0.33 & 0.08 & 0.26 & 0.24 & 0.07& 0.18 \\
\hline $IceCube$ & 0.61 &0.15  &0.48 &0.44 &0.13 &0.32 \\
\hline $\gamma-ray$ & 1.49 &0.36 &1.17 &1.09 &0.32 &0.79 \\
\hline
\end{tabular*} }
\end{center}
\caption{\footnotesize Separation in terms of $N_\sigma$ between 
the number of events 
obtained for different pairs of $\delta$ cases 
and the three $\nu$ flux limits, using the spherical geometry 
and $\Delta t_{obs}=$ 10days.} \label{sigma2}
\end{table}

\begin{table}[p]
\begin{center} {\footnotesize
\begin{tabular*}{0.75\textwidth}{@{\extracolsep{\fill}} |c!{\vrule width 1.5pt}cccccc|}
\hline
\multirow{2}{2.3cm}{~~$\nu$ flux limit}& \multicolumn{6}{c|}{Without $\delta$ distribution} \\
   &\multicolumn{1}{c}{EQ-IC} &
\multicolumn{1}{c}{EQ-Var} & \multicolumn{1}{c}{EQ-Min} &
\multicolumn{1}{c}{IC-Var}& \multicolumn{1}{c}{IC-Min} &
\multicolumn{1}{c|}{Var-Min} \\
\hline \hline$WB$ & 0.04&0.03 &0.05 &0.01 &0.01 &0.02  \\
\hline $IceCube$  & 0.07& 0.05&0.09 &0.02 &0.02 &0.04 \\
\hline $\gamma-ray$ & 0.18&0.13 &0.23 &0.04 &0.05 &0.09  \\
\hline\hline
& \multicolumn{6}{c|}{Using $\delta$ distribution} \\
  \hline \hline$WB$ &0.09 &0.16 &0.01 &0.06 &0.10 &0.16 \\
\hline $IceCube$  &0.17 &0.29 &0.02 &0.11 &0.19 &0.30 \\
\hline $\gamma-ray$ &0.43 &0.70 &0.04 &0.27 &0.47 &0.73   \\
\hline
\end{tabular*} }
\end{center}
\caption{\footnotesize Separation in terms of $N_\sigma$ between 
the number of events 
obtained for different pairs of $\delta$ cases 
and the three $\nu$ flux limits, using the cylindrical geometry 
and $\Delta t_{obs}=$ 10days.} \label{sigma3}
\end{table}

%%%%%%%%%%%%%%%%%%%%%%%%%%%%%%%%%%%%%%%%%%%%%%%%%%%%%%%%%%%%%%%%%%%%%%%
%%%%%%%%%%%%%% End of Manuscript
%%%%%%%%%%%%%%%%%%%%%%%%%%%%%%%%%%%%%%%%%%%%%%%%%%%%%%%%%%%%%%%%%%%%%%%

\end{document}